\documentclass{aastex631}

\newif\ifrev
\revfalse

\newif\ifrevv
\revvfalse

\newif\ifrevvv
\revvvfalse

\newcommand{\rev}[1]{\ifrev\leavevmode{\bf#1}\else #1\fi}
\newcommand{\revv}[1]{\ifrevv\leavevmode{\bf#1}\else #1\fi}
\newcommand{\revvv}[1]{\ifrevvv\leavevmode{\bf#1}\else #1\fi}

\begin{document}

\title{Asteroseismic Structure Inversions of Main-Sequence Solar-like Oscillators with Convective Cores}

\correspondingauthor{Lynn Buchele}
\email{lynn.buchele@h-its.org} 

\author[0000-0003-1666-4787]{Lynn Buchele} 
\affiliation{Heidelberg Institute for Theoretical Studies, Schloss-Wolfsbrunnenweg 35, 69118 Heidelberg, Germany}
\affiliation{Center for Astronomy (ZAH/LSW), Heidelberg University, Königstuhl 12, 69117 Heidelberg, Germany}

\author[0000-0003-4456-4863]{Earl P.~Bellinger}
\affiliation{Department of Astronomy, Yale University, PO Box 208181, New Haven, CT 06520-8101, USA}

\author[0000-0002-1463-726X]{Saskia Hekker}
\affiliation{Heidelberg Institute for Theoretical Studies, Schloss-Wolfsbrunnenweg 35, 69118 Heidelberg, Germany}
\affiliation{Center for Astronomy (ZAH/LSW), Heidelberg University, Königstuhl 12, 69117 Heidelberg, Germany}

\author[0000-0002-6163-3472]{Sarbani Basu}
\affiliation{Department of Astronomy, Yale University, PO Box 208181, New Haven, CT 06520-8101, USA}

\begin{abstract}

Asteroseismic inferences of main-sequence solar-like oscillators often rely on best-fit models. However, these models cannot fully reproduce the observed mode frequencies, suggesting that the internal structure of the model does not fully match that of the star. Asteroseismic structure inversions provide a way to test the interior of our stellar models. Recently, structure inversion techniques were used to study 12 stars with radiative cores. In this work, we extend that analysis to 43 main-sequence stars with convective cores observed by {\it Kepler} to look for differences in the sound speed profiles in the inner 30\% of the star by radius. For around half of our stars, the structure inversions show that our models reproduce the internal structure of the star, where the inversions are sensitive, within the observational uncertainties.  For the stars where our inversions reveal significant differences, we find cases where our model sound speed is too high and cases where our model sound speed is too low. We use the star with the most significant differences to explore several changes to the physics of our model in an attempt to resolve the inferred differences. These changes include using a different overshoot prescription and including the effects of diffusion, gravitational settling, \rev{and radiative levitation.} We find that the resulting changes to the model structure are too small to resolve the differences shown in our inversions.
\end{abstract}

\section{Introduction} \label{sec:intro}
Among the stars observed by \emph{Kepler}, high-precision oscillation mode frequencies have been determined for around 100 main-sequence solar-like oscillators \citep{ 2016MNRAS.456.2183D, 2017ApJ...835..172L}. This sample has been used to study a variety of physical processes including chemical transport \citep{2018MNRAS.477.5052N, 2018A&A...618A..10D,2019MNRAS.489.1850V,2022A&A...666A..43M,2024A&A...684A.113M}, convection in stellar cores \citep{2020MNRAS.493.4987A, 2020MNRAS.497.4042Z, 2023A&A...676A..70N}, rotation \citep{2023A&A...673L..11B}, and magnetic fields \citep{2018ApJS..237...17S,2018A&A...611A..84S,2020MNRAS.496.4593K}. This work often involves finding a best-fit model for each star using a stellar evolution code. Best-fit models are generally found by matching the observed frequencies of a star or by fitting parameters derived from those frequencies, such as the frequency separation ratios \citep{2003A&A...411..215R} or glitch signatures due to helium ionization \citep{2017ApJ...837...47V}, while matching the position of the star on the HR diagram. In general, however, these models are unable to fully reproduce the observed parameters, suggesting that there are still some deficits in our understanding of stellar interiors. 

Fortunately, the large number of precise oscillation modes observed in these stars makes it possible to take the analysis further using structure inversions. This technique, developed for geology \citep{1968GeoJ...16..169B} and used extensively in helioseismology \citep[for a review see, for example,][]{2016LRSP...13....2B, 2021LRSP...18....2C}, uses the inherent sensitivity of each oscillation mode to infer differences between the interior structure of a star and a given best-fit model (see e.g. \citealt{1991sia..book..519G, 1993afd..conf..399G, 2006mha..book.....P, 2020ASSP...57..171B, 2022FrASS...9.2373B}). These inferred differences can be used to test how well the interior structure of our models matches that of observed stars, as well as provide information on what changes may be necessary to improve our models.  

\begin{figure} 
    \epsscale{0.75}
    \plotone{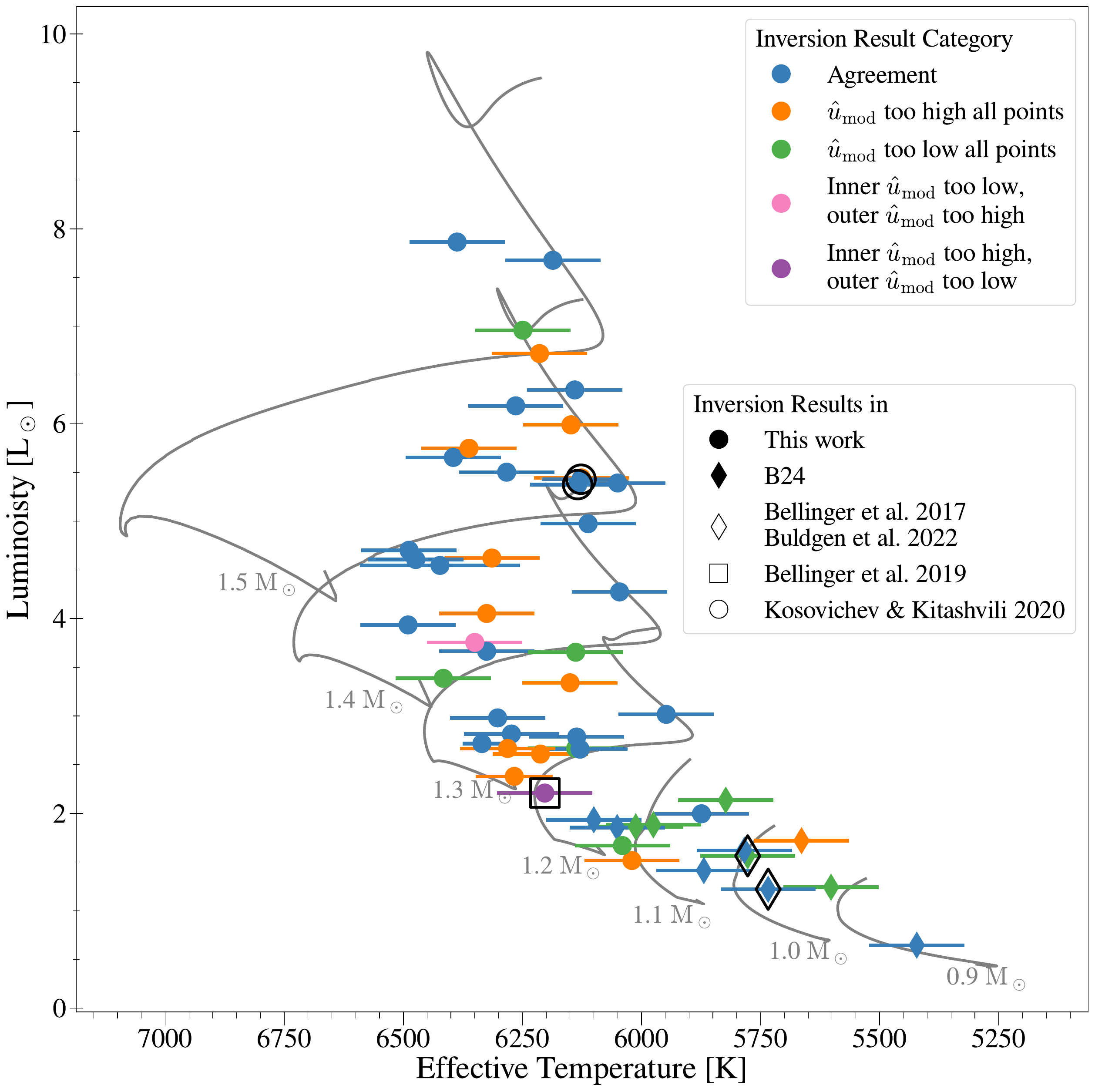}
    \caption{Hertzsprung–Russell diagram for main-sequence solar-like oscillators with inversion results available. Stars have been categorized based on their inversion results in this work and in B24, represented by the color of the symbol. Stars with other inversion results available are indicated with larger open symbols. The uncertainties of the luminosity values are smaller than the points. Stellar evolutionary tracks of several masses are shown for reference.}
    \label{fig:HRD}

\end{figure}

In Figure~\ref{fig:HRD}, we show the existing sample of main-sequence solar-like oscillators studied using asteroseismic structure inversions. \citet{2024ApJ...961..198B}, henceforth B24, presented results for 12 stars with radiative cores, including the solar analogues 16~Cyg~A and B which were also studied by \citet{2017ApJ...851...80B} and \citet{2022A&A...661A.143B}. Structure inversions have also been used to study a main-sequence star with a small convective core \citep{2019ApJ...885..143B} and two stars evolved enough to exhibit mixed modes \citep{2020IAUS..354..107K}. All three of these stars are in the sample presented here, where we extend the work of B24 to cover main-sequence solar-like oscillators with convective cores observed by \emph{Kepler}. \revvv{Our primary goals of this work are to determine which stars have observational data suitable for inversions and to examine how well models created using common modeling choices reproduce the structure of observed stars.}

\section{Forward Modeling} 
\label{Section-mods}
As structure inversions infer differences between a star and a model, the first step is to find a suitably close reference model, typically the best-fit model from a grid-based modeling or optimization procedure. The process of finding such a model is called forward modeling. To find our reference models, we used a grid-based method similar to that used by B24. We constructed \rev{a grid of 24,530 tracks} using the stellar evolution code MESA \citep{Paxton2011, Paxton2013, Paxton2015, Paxton2018, Paxton2019, Jermyn2023}. The details of the grid that are unchanged between this work and B24 are provided in Appendix~\ref{Appendix-grids}. There are two important changes which we discuss here. It is well known that including diffusion and gravitational settling of elements without also including the effects of radiative levitation produces models with unrealistic surface abundances in this mass range \citep[e.g.][]{2015ads..book.....M, 2018A&A...618A..10D}. However, including radiative levitation significantly increases the computation time of models, such that it would be difficult to compute the number of models necessary to cover the parameters space of the observations. We chose to compromise and evolve our tracks without including diffusion, settling, or radiative levitation. Additionally, since we are dealing with stars with convective cores, we use the exponential overshooting \citep{1996A&A...313..497F, 2000A&A...360..952H} scheme implemented in MESA where the overshoot region is treated as fully mixed without changing the thermal gradient. This is described in detail in \citet{Paxton2011}. The parameters varied in this grid are mass, initial helium abundance, initial metallicity, mixing length parameter, and overshooting parameter. To cover the parameter space efficiently, we varied each parameter using a Sobol sequence  \citep[see Appendix~B of][]{2016ApJ...830...31B, SOBOL196786} within the ranges listed in Table \ref{tab:om_grid_param}. For each model in the grid, we calculated the adiabatic frequencies using GYRE \citep{Townsend2013}. We then scanned the grid to find the model parameters that best fit the frequencies, effective temperature, and metallicity by minimizing: 
\begin{eqnarray}
\chi^2_{\rm{fit}} = \frac{\left(T_{\rm{eff,obs}} - T_{\rm{eff,mod}}\right)^2}{\sigma^2_{T_{\rm{eff}}}} \\
+ \frac{\left(\rm{[Fe/H]}_{\rm{obs}} - \rm{[Fe/H]}_{\rm{mod}} \right)^2}{\sigma^2_{\rm{[Fe/H]}}} \\
+ \frac{1}{N} \sum_{i}^{N} \frac{\left(\nu_{i, \rm{obs}} - \nu_{i, \rm{mod}}\right)^2}{\sigma^2_{\nu, i}},
\label{equ:chi2_fit} 
\end{eqnarray}
where \(N\) is the number of observed frequencies, \(\nu_{i}\) is the frequency that corresponds to the \(i\)th  pair of radial order ($n$) and spherical degree ($l$) where the model's frequencies have been corrected for surface effects using the two-term correction from \citet{2014A&A...568A.123B},  \(\sigma\) denotes the uncertainty of the observed parameter, and the subscripts `obs' and `mod' denote the observations and the model, respectively. Our definition of $\chi^2_{\rm{fit}}$ treats all the frequencies as a single observation with the same weight as each spectroscopic observation. This choice is common in asteroseismic modeling pipelines (see, for example, the ASTFIT pipeline described in  \citealt{2015MNRAS.452.2127S}). Each mode can be treated as an independent observation by removing the factor of $1/N$. In synthetic tests, \citet{2021MNRAS.508.5864C} find that this weighting recovers the correct stellar parameters only when the physics of the grid matches the physics of the synthetic star exactly. As we perform structure inversions in order to determine if the physics in our models accurately represents what we observe, we therefore opt \rev{to treat all frequencies as a single observation}.

While scanning the grid, we interpolated along each track, but not between tracks. This is the same method as that of B24 with one change --- in this work we interpolated in age instead of central hydrogen abundance, as the central hydrogen abundance does not decrease monotonically in stars where a convective core emerges after the zero-age main-sequence. \rev{From this procedure, we obtain the best-fit parameters, which are then used to calculate the reference model of the structure inversions.} Our values of \(T_{\rm{eff}}\) and [Fe/H] come from \citet{Furlan_2018, Mathur_2017, Morel_2021}, with the specific source for each star given in Appendix~\ref{Appendix-mods}. We also provide, in Appendix~\ref{Appendix-mods}, the parameters of our best-fit model, \revvv{a comparison to the parameters reported in \citet{2017ApJ...835..173S}, and a comparison of our $\chi^2_{\rm{fit}}$ distribution to that obtained using the YMCM pipeline of \citet{2017ApJ...835..173S}}. Additionally, we provide both the FGONG structure file and the inlist used to generate each model at \url{https://zenodo.org/records/15341350}. 

\begin{deluxetable}{lcc}

\tablecaption{Grid Parameters} 
\label{tab:om_grid_param} 
\tablehead{\colhead{Parameter} & \colhead{Minimum Value} & \colhead{Maximum Value}} 
\startdata
$M/\rm{M}_{\odot}$ & 1.1 &  1.7 \\
$Y_{\rm{initial}}$ & 0.24 &  0.4 \\
$Z_{\rm{initial}}$ & 0.0005  &  0.07 \\
$\alpha_{\rm{mlt}}$ & 1.3 & 2.4 \\
$f_{ov}$ & 0 & 0.08 \\
\enddata
\end{deluxetable}

\section{Structure Inversions}  

\rev{With a suitable reference model for each star in our sample, we now turn to the process of an asteroseismic structure inversion. Structure inversions use the frequency differences between a star and its best-fit model to infer the underlying structure differences. We chose to express the structure differences in terms of the dimensionless squared isothermal sound speed (\(\hat{u}\)) and helium mass fraction (\(Y\)). In terms of the more common structure variables of pressure \((P)\) and density \((\rho)\),
\begin{equation}
\hat{u} = \frac{P}{\rho} \: \frac{R}{M}
\end{equation}
where \(R\) and \(M\) are the stellar radius and mass, respectively. This choice of variables is well suited for asteroseismic targets \revvv{\citep{1993ASPC...40..541G, 2003Ap&SS.284..153B, 2020ASSP...57..171B}}, as the oscillations are mostly insensitive to \(Y\). This makes it easier to isolate the differences due to a change in \(\hat{u}\).}

\rev{Mathematically, the sensitivity of each mode frequency to a small change in the structure is expressed as:}
\begin{equation}
\frac{\delta \hat{\nu}_i}{\hat{\nu}_i} = \int K_{i}^{(\hat{u}, Y)} \frac{\delta \hat{u}}{\hat{u}} \,\textrm{d}r + \int K_{i}^{(Y,\hat{u})} \delta Y \,\textrm{d}r + \text{ higher order terms}.
\label{equ:mode_kernels}
\end{equation}
Such an equation can be written for each mode  \(i\), where the index $i$ of the mode again corresponds to a specific pair of \(n,l\). The relative frequency difference (\(\delta \hat{\nu}_i / \hat{ \nu}_i\)) is related to the structure differences between the model and the observed star through the mode kernel functions \(K_i\). These mode kernels are known functions of the reference model, found through a linear perturbation of the oscillation equations (for more details, see \citealt{1991sia..book..519G}, \citealt{1999JCoAM.109....1K}, or \citealt{2002ESASP.485...95T}). \revvv{There are two potential sources of frequency differences that are not accounted for in the mode kernels: surface effects and differences in mean density. We correct for the surface term during the calculation of our frequency differences using the same correction applied when finding our reference model, the two-term correction of \citet{2014A&A...568A.123B}. We account for mean density differences by  inverting for dimensionless structure variables. This requires us to use the difference in dimensionless frequency ($\hat{\nu}$).} These differences are calculated by scaling the dimensional frequency differences by the large frequency separation, $\Delta \nu$:
\begin{equation}
\frac{\delta \hat{\nu}_i}{\hat{\nu}_i} \approx \frac{\Delta \nu_{\rm{mod}}}{\Delta \nu_{\rm{obs}}} \frac{\nu_{i, \rm{obs}}}{\nu_{i,\rm{mod}}} - 1.
\label{equ:freq_dif}
\end{equation}
This works because $\Delta \nu$ carries the same dependence on the stellar mass and radius as the frequencies. The full derivation of Equation~\ref{equ:freq_dif} can be found in Appendix B2 of B24. \revvv{In this expression $\Delta \nu_{\rm{mod}}$ is calculated after applying the surface term correction.} \revvv{We note that both of these effects can be handled in different ways.} 

Alternative approaches to handling dimensional differences include using a different correction method \revv{\citep[e.g.][]{R98, 2003Ap&SS.284..153B, 2021ApJ...915..100B},  including the mean density in the fitting procedure  \citep[e.g.][]{2022A&A...661A.143B}, or adding a term to Equation \ref{equ:mode_kernels} \revvv{\citep[e.g.][]{1993ASPC...40..541G,2020IAUS..354..107K}}. Following the arguments outlined in Appendix B2 of B24, we expect that the inversion procedure will suppress the effects of differences in mean density regardless of the correction method used. \citet{2019ApJ...885..143B} shows this explicitly as inversions using models of different masses and radii return the same results. \revv{The surface term can be accounted for during the inversion by adding a term to Equation \ref{equ:mode_kernels} \citep[e.g.][]{1998ESASP.418..505R, 2016ApJ...830...31B,2022A&A...661A.143B}. We have tested this approach and found no difference in the final inversion results.}}

If the structure differences are known, then the right-hand side of Equation \ref{equ:mode_kernels} can be used to calculate the corresponding frequency differences. When comparing an observed star to its best-fit model, however, we  know the frequency differences and seek to infer the underlying structure differences. We accomplish this through the method of optimally localized averages \citep{1968GeoJ...16..169B, 1970RSPTA.266..123B}. This constructs a linear combination of mode kernels that localizes the overall sensitivity around a single target radius, \(r_0\). Neglecting higher-order effects, Equation \ref{equ:mode_kernels} becomes: 
\begin{equation}
\sum_i^N c_i \frac{\delta \hat{\nu}_i}{\hat{\nu}_i} = \int \mathcal{K}_{r_0} \frac{\delta \hat{u}}{\hat{u}} \,\textrm{d}r + \int \mathcal{C}_{r_0}\, \delta Y \,\textrm{d}r. 
\label{equ:comb_kernels}
\end{equation}
Here \revv{$N$ is the total number of modes}, \(c_i\) are known as the inversion coefficients, \(\mathcal{K} = \sum_i^N c_i K_i^{(\hat{u},Y)}\) is called the averaging kernel, and \(\mathcal{C} = \sum_i^N c_i K_i^{(Y,\hat{u})}\) is the cross-term kernel. 
When the inversion coefficients are chosen such that \(\mathcal{K}\) is localized around \(r_0\) and normalized to 1, and the amplitude of \(\mathcal{C}\) is small everywhere, then Equation \ref{equ:comb_kernels} reduces to
\begin{equation} 
\sum_i^N c_i \frac{\delta \hat{\nu}_i}{\hat{\nu}_i} \approx \int \mathcal{K}_{r_0} \frac{\delta \hat{u}}{\hat{u}} \: \textrm{d}r \approx \left< \frac{\delta \hat{u}}{\hat{u}} \right>_{r_0}. 
\label{equ:avg_kern} 
\end{equation} 
Thus once the inversion coefficients are known, the sum on the left-hand side provides a localized average of the difference in \(\hat{u}\) around \(r_0\). 

To find the inversion coefficients, we used the method of multiplicative optimally localized averages (MOLA), which constructs the averaging kernel by penalizing any amplitude away from the target radius. For details on the implementation of MOLA, see \citet[][Chapter~10]{2017asda.book.....B}. In this process, we must choose two trade-off parameters: \(\beta\), the cross-term suppression parameter, and \(\mu\), the error suppression parameter. We chose our parameter values using the same method as B24. Briefly, this method sets \(\beta = 0\) as the choice of \(Y\) as the second variable naturally suppresses the amplitude of the cross-term kernel. We then chose a value of \(\mu\) that correctly recovers the known values of \(\delta \hat{u} / \hat{u}\) between our reference model and a small set of calibration models. These models came from our grid and have slightly larger values of \(\chi^2_{\rm{fit}}\) than our reference model. 

For each target star, we attempted to construct an averaging kernel for six target radii: \(r_0/R = 0.05, 0.1, 0.15, 0.2,0.25,0.3\). In general, the presence of a convective core made it more difficult to localize sensitivity at target radii close to the boundary of the core, and so in most cases the innermost target radius we report is \(r_0/R = 0.15\). It is possible that frequencies derived from radial velocity measurements instead of photometric measurements could expand this range by providing more modes overall, which would help to suppress the sensitivity to the boundary of the convective core, and by providing more \(l=3\) modes which make it easier to localize averaging kernels at larger target radii. The uncertainties of our inversion results are calculated using a Monty Carlo simulation to account for possible error correlations introduced by our corrections for the mean density and surface effect (for the specific details, see B24). We also report the FWHM of each averaging kernel as a measure of the resolution of each inversion. 

For 11 stars, our models showed that the lowest order quadrupole modes were  mixed acoustic-buoyancy modes. We have found that current linear inversion techniques are not suitable for mixed-modes (Buchele et al., in preparation) and hence, while we accounted for these mixed modes when fitting our models, we removed these modes from the mode set used for the structure inversions. Table \ref{tab:mm_remove} shows the stars with mixed modes present and how many quadrupole modes were excluded from our inversions.

\begin{deluxetable}{lc}
\tablecaption{Stars with Mixed Modes Removed} 
\label{tab:mm_remove} 
\tablehead{\colhead{KIC Number} & \colhead{Number of \(l=2\) modes removed} } 

\startdata
8228742 & 3 \\
7940546 & 3 \\ 
10068307& 4 \\ 
12317678& 1 \\ 
3632418& 1 \\
10162436& 1 \\ 
9353712& 1 \\
9414417& 1 \\
3456181& 1 \\
12069127& 2\\
6679371& 2 \\
\enddata

\end{deluxetable} 

\section{Results} 
\label{sect:results}
\begin{figure}
    \epsscale{1.2}
    \plotone{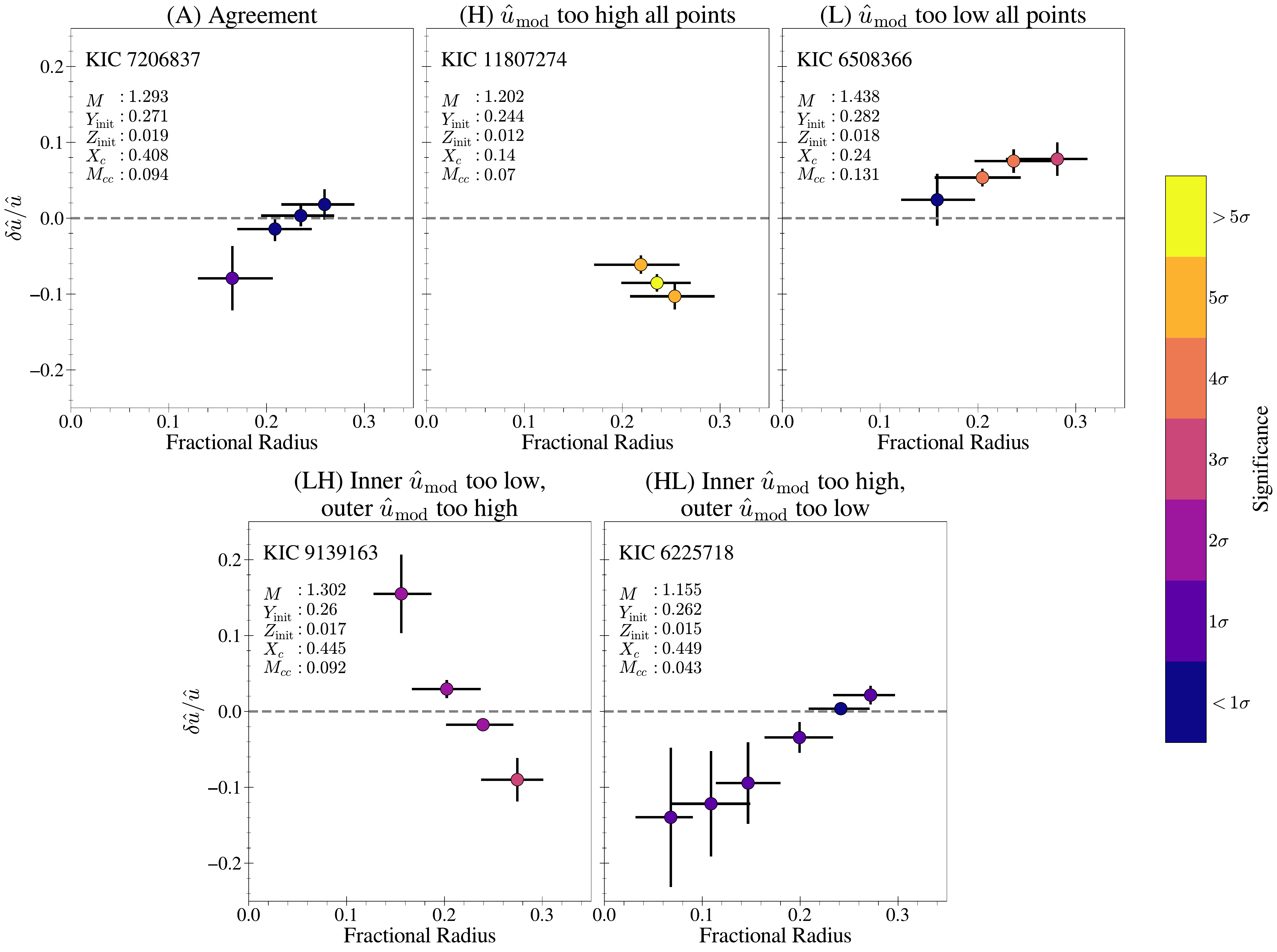}
    \caption{Inversion results for one star in each category. Each shows the relative differences in \(\hat{u}\) between observation and best-fit model inferred by the inversions, in the sense of (Star $-$ Model)/Model. The vertical error bars indicate the uncertainty of each inversion result from the propagation of the uncertainty of the observed frequencies. The horizontal error bars represent the FWHM of the averaging kernel. The dashed horizontal line indicates complete agreement between the model and observations; points above this line imply that \(\hat{u}\) of the star is larger than that of the model. The color bar indicates the statistical significance of the inferred difference, with lighter colors showing more significant results. We also provide the mass (\(M\), in \(M_\odot\)), initial helium mass fraction (\(Y_{\rm{init}}\)), initial metallicity (\(Z_{\rm{init}}\)), central hydrogen mass fraction (\(X_c\)), and mass of the convective fore (\(M_{cc}\), in \(M_\odot\)) of each model.}
    \label{fig:category} 
\end{figure}

We divide our 43 stars into five categories based on their inversion results: (A) stars with no significant disagreement in the region probed by inversions, (H) stars for which all significant differences show that the model \(\hat{u}\) is too high, (L) stars for which all significant differences show the model \(\hat{u}\) is too low, (LH) stars where the model \(\hat{u}\) is too low in the center and too high in the outer points probed by the inversions, and (HL) stars where the model \(\hat{u}\) is too high in the center and too low in the outer points probed by the inversions. In Figure \ref{fig:category}, we show an example of inversion results from one star in each category. 

Around half (24) of the stars fall into category (A). These models still show significant differences in the oscillation frequencies, even after correcting for the surface term, which suggests that the structure differences are either smaller than the observational uncertainties at the resolution given by the structure inversions or that the structure differences are at a location unable to be probed by the inversions. 
 Of the stars showing significant disagreement, 11 are in category (H), 6 are in category (L), and 1 each is in categories (LH) and (HL). Using the \(\chi^2_{\rm{inv}}\) parameter defined in B24, we search for correlations with a variety of model parameters and observations. In contrast to the earlier work, we find no significant correlations. One problem with the \(\chi^2_{\rm{inv}}\) metric is that it only measures the significance of the inversion results, not whether the differences inferred are positive or negative. To account for this, we also look for correlations between the model parameters and \(a \chi^2_{\rm{inv}}\) where \(a = -1 (+1)\) for stars where the most significant inferred difference is negative (positive). We also find no significant correlations in this case.

For 13 stars, we find models with both convective and radiative cores in our calibration set. In general, the models of the stars that do have convective cores have small ones, implying the structure differences between the calibration models are relatively small. The distribution of the inversion results within this subsample is similar to that of the whole sample, suggesting that the differences we infer are not due to the ambiguity in whether the core is convective or radiative.

We discuss here only a few of our 43 stars, focusing on the stars that other works have also analyzed with structure inversions. We present the full inversion results for each star in Appendix \ref{Appendix-inv}. 

\begin{figure}
    \epsscale{0.7}
    \plotone{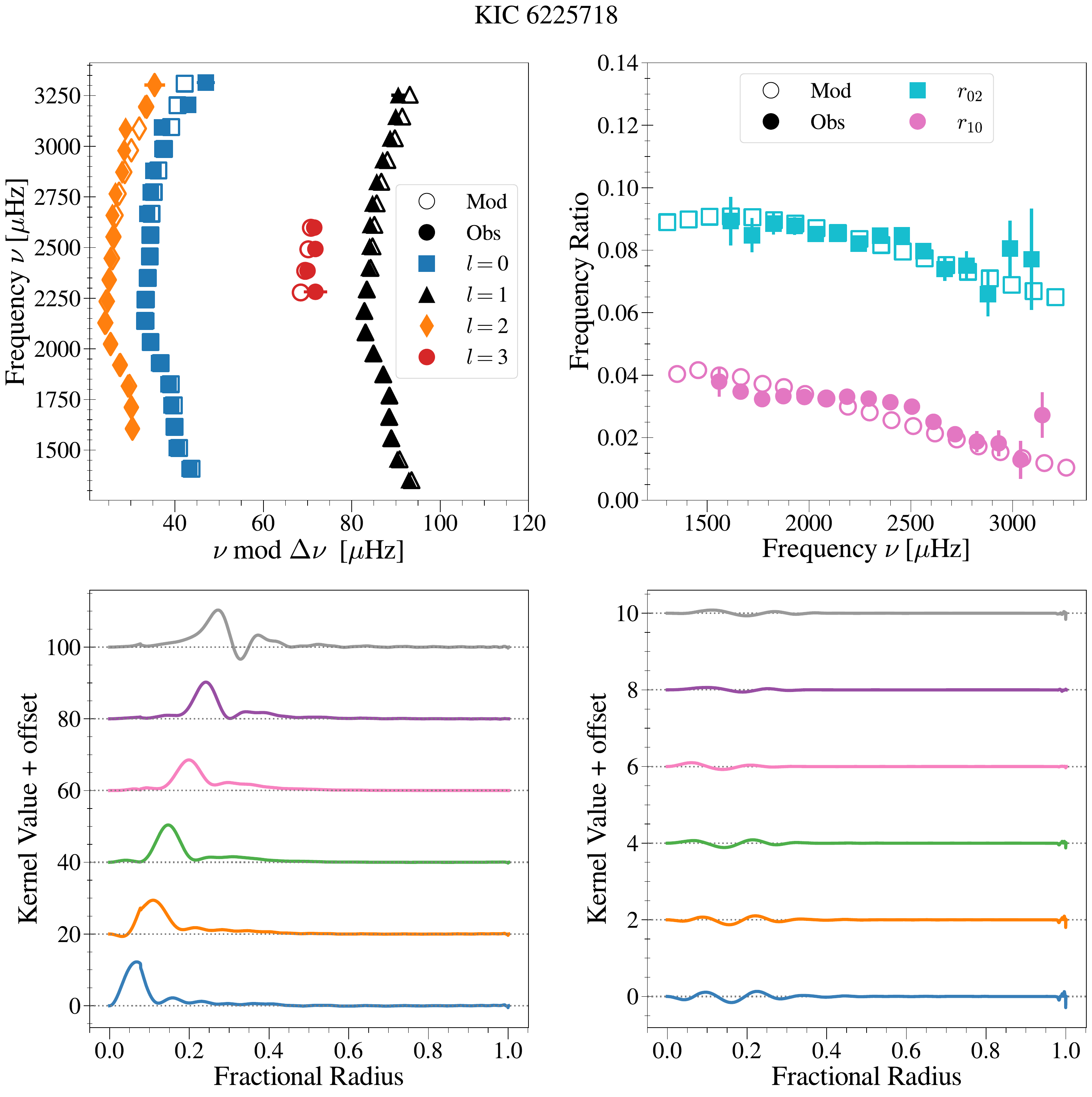}
    \caption{Information about the best-fit model of KIC~6225718. The top left plot shows the frequency \'echelle diagram comparing the frequencies of the reference model to the observations after correcting for surface effects. The top right plot shows the frequency separation ratios of the reference model and observations. The bottom left plot shows the averaging kernels and the bottom right the cross-term kernels. Note that the $y$-axis scale differs \rev{by an order of magnitude} between the two plots on the bottom row.}
    \label{fig:6225718} 
\end{figure}

\subsection{\rev{KIC~6225718}}
\begin{figure} 
\epsscale{0.8}
\plotone{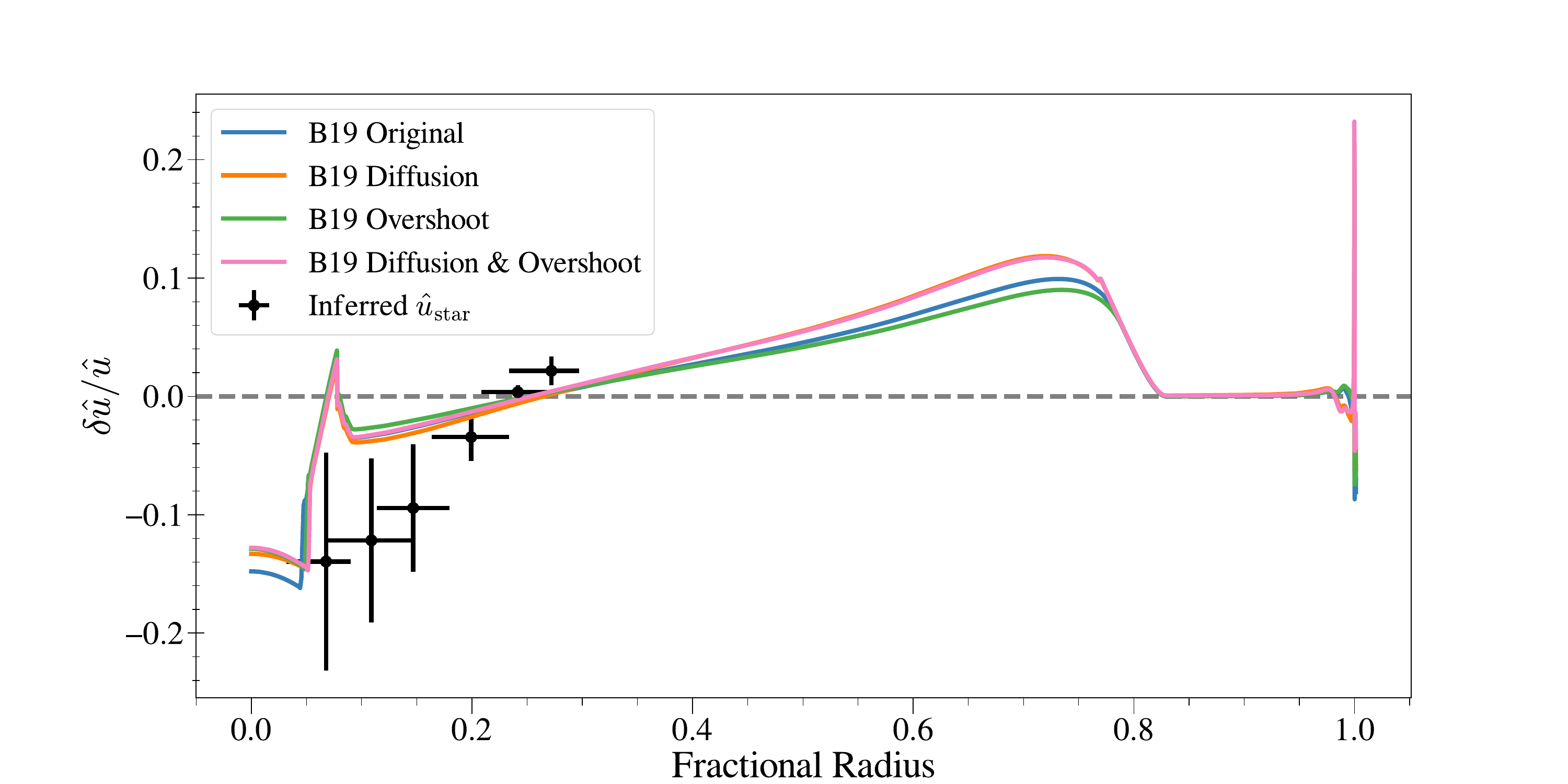} 
\caption{\revvv{Structure differences between the reference model for KIC~6225718 used in this work and those used in \citet{2019ApJ...885..143B} (B19). Each line shows the relative difference in \(\hat{u}\) between our reference model and the model from \citet{2019ApJ...885..143B} constructed using the modeling physics indicated in the legend.  We indicate 0 with a dashed horizontal line \revv{and plot our inversion results for this star in black points}
}} 
\label{fig:6225718_struct_comp} 
\end{figure}

The first star we discuss in detail is KIC~6225718.  Our inversion results for this star are shown in Figure \ref{fig:category}. We show in Figure \ref{fig:6225718} the frequency  \'echelle diagram and frequency separation ratios of our best-fit model, as well as, our averaging and cross-term kernels. This star has already been studied using structure inversions by \citet{2019ApJ...885..143B} which allows us to compare our results. 
\revvv{
Structure inversions infer differences relative to a given reference model using a given set of averaging kernels. As such,  any comparison of different inversion results must be considered in the context of the reference model and averaging kernels used. The averaging kernels we use are very similar to those used in \citet{2019ApJ...885..143B}. We focus on the structure for our comparisons. We have obtained the reference models used in \citet{2019ApJ...885..143B} to test the variation of the modeling physics and compare them to our reference model in Figure~\ref{fig:6225718_struct_comp}.  In all cases, the models of \citet{2019ApJ...885..143B} are more similar to each other than to our reference model, although the model with physics closest to our choices (overshoot without diffusion) is the closest to our reference model. This suggests that the differences between our models are due to differences in the fitting procedure.}

\revvv{Both works find that the model \(\hat{u}\) is too low in the outermost regions probed by inversions and too high in the innermost regions, with the crossover occurring around \(r/R \approx 0.25\). \citet{2019ApJ...885..143B} find a maximum difference at \(r/R \approx 0.1\), while our maximum difference is found around \(r/R \approx 0.05\). This difference in inversion results is explained by examining the $\hat{u}$ profiles of the various reference models. Our model has a higher $\hat{u}$  in the center of the model. Due to the width of the averaging kernel these structure differences still influence the inversion results of the lowest target radii, despite being located below the lowest target radius.  
In all cases, the structure differences between the models are smaller than the 1$\sigma$ inversion result uncertainty.}


\subsection{\rev{KIC~10162436 and KIC~5773345}}
\begin{figure}
    \gridline{\fig{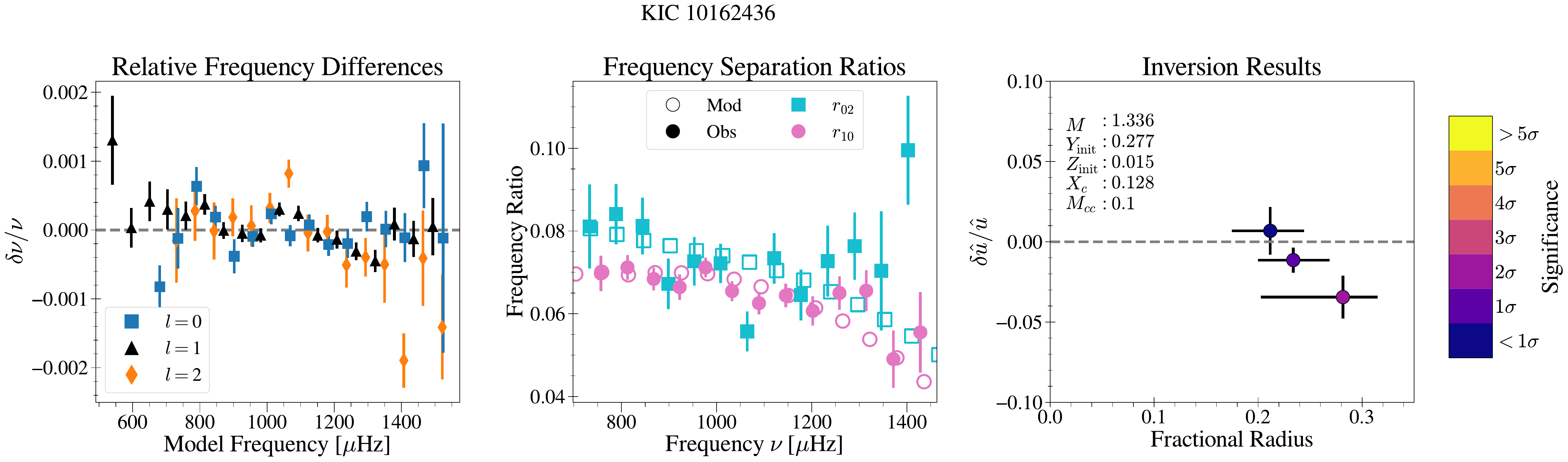}{\textwidth}{}}
    \gridline{\fig{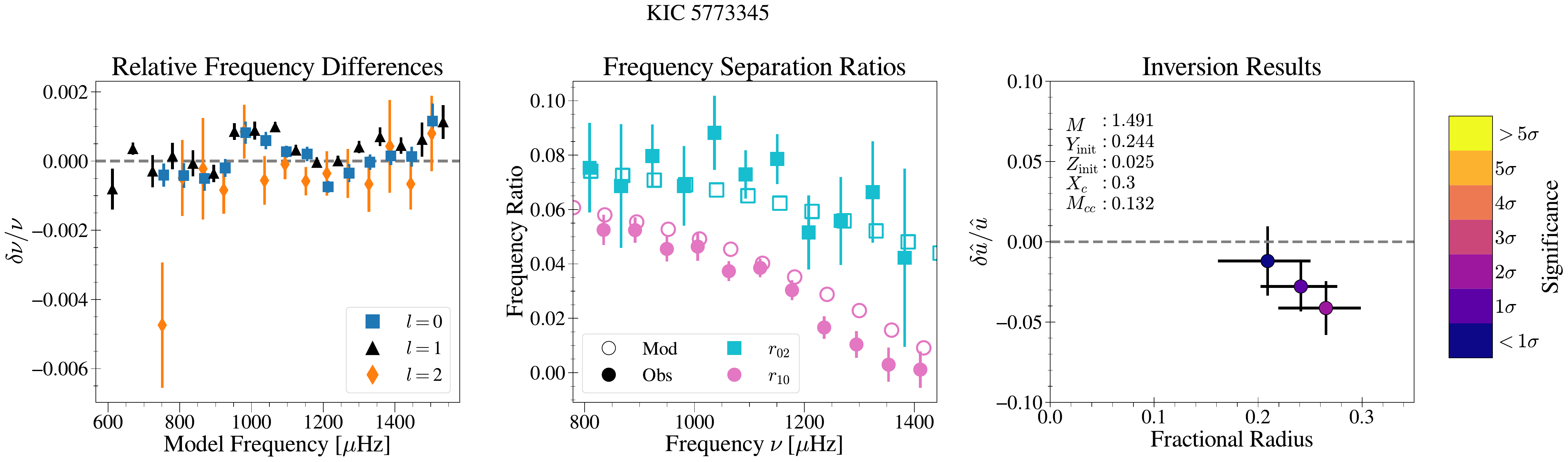}{\textwidth}{}}
    \caption{Frequency differences, \rev{frequency separation ratios}, and inversion results for our models of the two stars shown in \citet{2020IAUS..354..107K}. The top row shows the results for KIC~10162436. The figure on the left shows the relative frequency differences, after correcting for surface effects. \rev{The center panel shows the frequency separation ratios  of the observed star and our reference model.} The right panel shows the inversion results, where all symbols and colors have the same meaning as in Figure \ref{fig:category}. The lower row shows the same information for KIC~5773345. }
    \label{fig:K&Kstars} 
\end{figure}

\citet{2020IAUS..354..107K} present inversion results for two stars that are also in our sample: KIC~10162436 and KIC~5773345. Our results for these stars are shown in Figure \ref{fig:K&Kstars}. \revvv{Both of these stars, as well as the star discussed in the next section, show glitches in their frequency separation ratios. These go beyond the change of slope expected due to core overshooting \citep{2016A&A...589A..93D}.  Previous works \citep{2012A&A...544L..13L, 2023A&A...673A..49D} have shown that similar glitches can be induced in the frequency separation ratios of models by including large amounts (1-2 times the pressure scale hight) of convective penetration at the base of the convection zone, although this is considered physically unrealistic. The inability of models to reproduce these glitches in the ratios is an additional suggestion that something is missing in our stellar models. } 

Directly comparing our inversion results to theirs is slightly more difficult than with \cite{2019ApJ...885..143B}. For both stars, they find mixed modes with \(l = 1\) and \(l=2\). Our model of KIC~10162436 has one mixed \(l=2\) mode, which we exclude from our inversions, and no \(l=1\) mixed mode. In the case of KIC~5773345, our model shows no mixed modes at all. In addition, the frequency differences, even of pure acoustic modes, between their models and the observations are significantly larger than ours. \revv{We attribute these differences to differences in the modeling procedure. 
\citet{2020IAUS..354..107K} use the parameters from the YMCM modeling pipeline presented in \citet{2017ApJ...835..173S}, including the mixing length parameter and stellar age, to compute a model using MESA. However, the YMCM models were computed using YREC, a different stellar evolution code, and care must be taken when using best fit parameters, especially mixing length parameter and stellar age, across different codes to ensure that the implemented physics matches as closely as possible. In particular, the nuclear reaction rates and formulation of mixing length theory differ between YREC and the defaults used in MESA. It is unclear whether the authors of \citet{2020IAUS..354..107K} made the necessary changes to MESA to match the original YREC configuration. These differences likely explain the large frequency differences that \citet{2020IAUS..354..107K} find between the observed and modeled frequencies, despite the model parameters being the same. Taken together, these differences} suggest that the structure of the reference models used in \citet{2020IAUS..354..107K} and this work are different. Additionally, while \citet{2020IAUS..354..107K} do not show their averaging kernels, the spread indicated by their horizontal error bars are much wider than ours. Thus we cannot directly compare our inversion results to those given in \citet{2020IAUS..354..107K}. Nevertheless, we note that we infer differences in \(\hat{u}\) of similar magnitude for both KIC~10162436 and KIC~5773345.

\subsection{\rev{KIC~11807274}}
\begin{figure} 
\epsscale{0.8}
\plotone{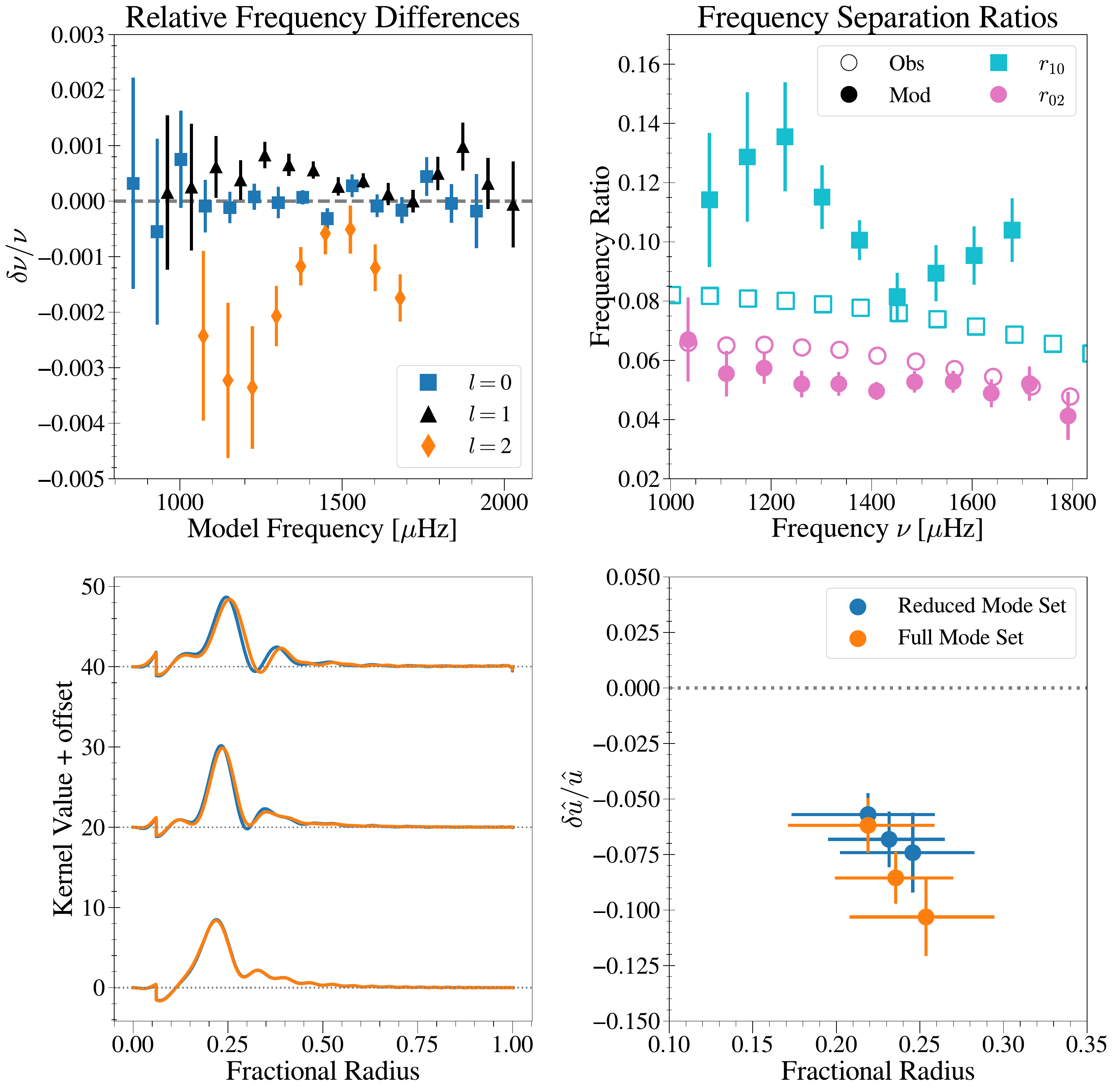} 
\caption{\rev{
Results of the modeling and modeset testing of KIC~11807274. The top-left figure shows the relative frequency differences, after correcting for surface effects. The top-right figure shows the frequency separation ratios of the observed star and our reference model. The bottom right figure shows the original averaging kernels and those constructed when excluding the three lowest frequencies $l=2$ modes. The bottom left figure shows the results of inversions using each mode set. 
}} 
\label{fig:1180_details} 
\end{figure}

\rev{We now turn to the star with the most significant differences inferred by our inversions, KIC~11807274. We show in Figure~\ref{fig:1180_details} the frequency differences and frequency separation ratios for this star. \revvv{This star also shows the glitches in the ratios discussed in the previous section.}  Our reference model is in full agreement with the observed values of $T_{\rm{eff}}$ and [Fe/H]. The largest frequency differences are seen in the lower order quadrupole frequencies, which exhibit a glitch structure not reproduced in any model. To understand how sensitive our results are to these discrepant frequencies, we repeat both our modeling and inversion procedure excluding the lowest three quadrupole models. Our fitting procedure results in the same model as we found using the entire mode set. We show the averaging kernels and inversion results of the reduced mode set compared with the full mode set in Figure~\ref{fig:1180_details}. Removing these modes results in slightly different averaging kernels, most notably for the highest target radius, where the maximum of kernel amplitude is shifted towards the center of the star. The differences inferred with these new averaging kernels are smaller than with the full mode set, however significant differences remain.}

\subsection{Changes in Input Physics} 
\begin{figure}
    \epsscale{1}
    \plotone{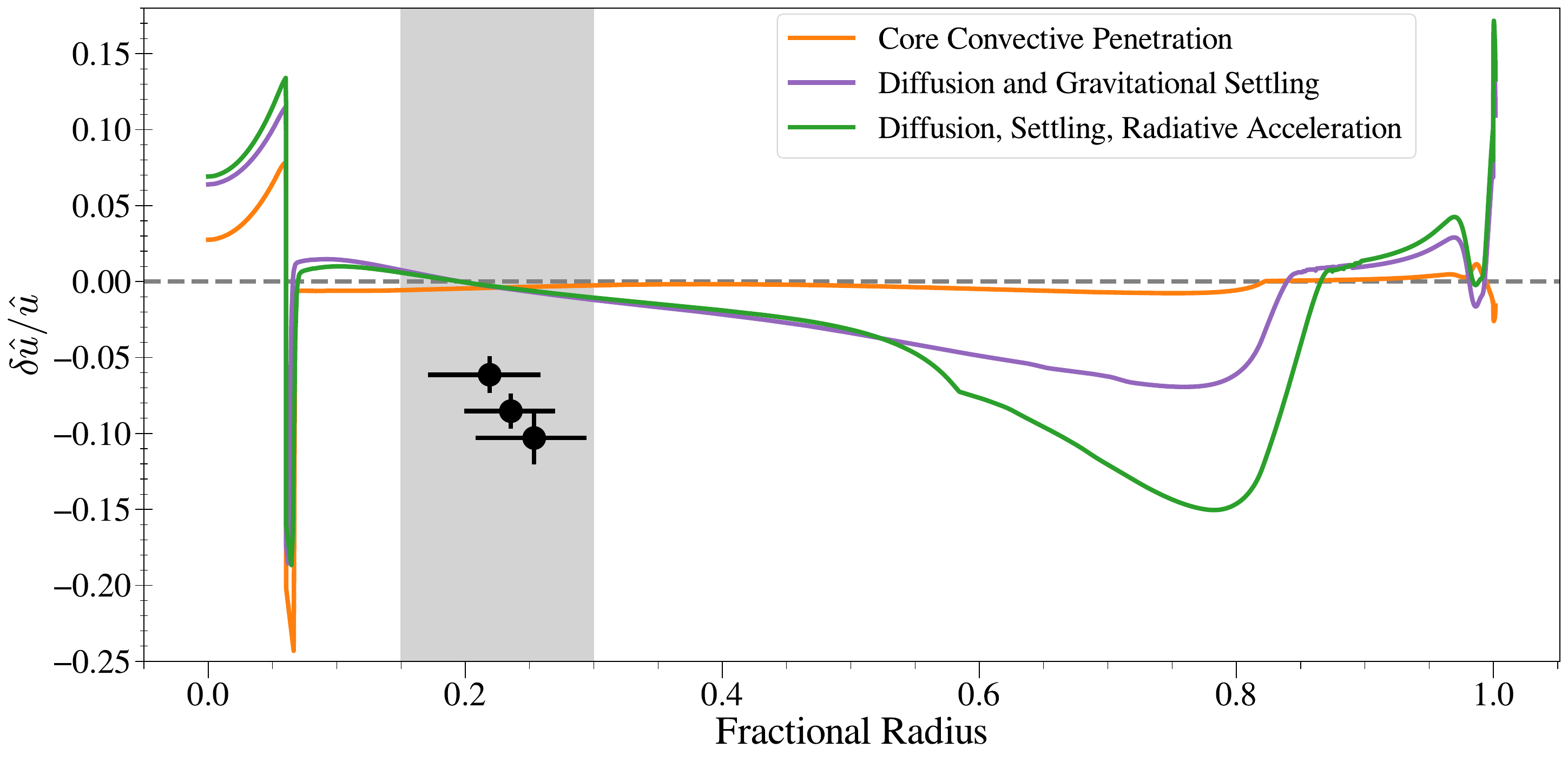}
    \caption{Results of varying the several mixing processes for models of KIC~11807274. Each line shows the relative difference in \(\hat{u}\) between our original reference model and a model with different physics. We indicate 0 with a dashed horizontal line \revv{and plot our inversion results for this star in black points}. The vertical shading indicates the fractional radii where our inversions are sensitive. In all three cases, the spike around $r/R \approx 0.1$ is due to differences in the boundary of the convective core. The larger changes in the \revv{ $\hat{u}$ profiles around $r/R \approx 0.8$} are due to differences in the depth of the outer convection zone. 
    }
    \label{fig:11807274-mixing_changes} 
\end{figure}

\revv{The fact that we find significant differences in many stars suggests that the physics commonly used in stellar modeling codes may need to be modified. A full study seeking to prove the accuracy of one set of physical choices over another will require extensive modeling of all stars for which structure inversions can be used. This modeling effort should test as many changes as possible to the modeling physics and explore the effect of fitting to different observables, such as the frequency separation ratios. While we think that such work is important to continue improving stellar models, it is beyond the scope of this work. Instead, here we present a few simple tests as examples of the changes to modeling physics that could be studied with inversions. For this we use the star where our inversions infer the most significant differences, KIC~11807274.}
In Figure~\ref{fig:11807274-mixing_changes}, we show the changes to the \(\hat{u}\) profile that result from three different changes to the physics. The first change we present is a change to core boundary mixing. Instead of calculating the overshoot using exponential overmixing, where only the composition of the overshooting region is changed, this model uses a step convective penetration \citep{1991A&A...252..179Z} scheme described in Appendix~\ref{Appendix-grids}, where both the composition and the temperature gradient are changed in the overshoot region. For this change, we find new model parameters from a new grid created with the changed overshooting scheme. This change results in a slightly larger convective core, which causes the spike in the relative difference around \(r/R \approx 0.05\). Otherwise, the main difference in the \(\hat{u}\) profile is within the convective core, below the radius where our inversions probe. In the region where our inversions are sensitive, the change to \(\hat{u}\) is in the correct direction according to our inversion results, but it is far too small to resolve the differences. In fact, the change is smaller than the uncertainties of our inversions. \rev{Several works \citep{ 2012A&A...544L..13L, 2023A&A...673A..49D} have suggested that a large amount of convective penetration at the base of the outer convection zone may explain glitches observed in the frequency separation ratios of F-type stars. We have tested this prescription as well and find a change to the \(\hat{u}\) at the base of the convective envelope \(r/R \sim 0.8\), but at the radius where our inversions are sensitive the structure is very similar to the model without this additional mixing. Hence, this change is  unable to resolve the differences inferred by our inversions.} 

\rev{The other changes we examined deal with the transport of chemical elements. We tested the effects of including element diffusion and gravitational setting only as well as accounting for diffusion, settling, and radiative levitation. \revv{In contrast to our test of convective penetration,} for both of these models we kept the same overshoot implementation and initial parameters \revv{(mass, composition, overshoot and mixing length parameters)} as our original reference model. \revv{The age of our new model is allowed to differ from the age of the original model. We choose the age along our new track which best fits the observations.} In the model including only diffusion and settling we used the inlist parameters of the }\verb|diffusion_smoothness| \rev{test suite case in MESA. In the model including diffusion, settling, and radiative levitation we adopt the MESA settings of the A0 model in  \citet{2022A&A...659A.162C}. In both of the new models we find the largest differences around the base of the convection zone where the transport processes have made the convection zone deeper.}
In the regions probed by our inversions, however, the changes are small enough to be within the uncertainties of our inversion results. 
\revv{Our choice to keep the initial parameters of the models constant between the different chemical mixing prescriptions represents the simplest possible test. In a full work seeking to fully resolve the inferred differences, these parameters should be inferred from a full grid as this change of physics is known to change the inferred mass, radius, and age of the star \citep[e.g.][]{2020A&A...633A..23D,2022A&A...666A..43M,2024A&A...684A.113M}, although how these changes affect the internal structure is not discussed in these works. For our purposes here, we seek only to provide some examples of the types of physical changes that can potentially be tested with structure inversions. }

\section{Conclusions}
\rev{In this work, we have extended the analysis from \citet{2024ApJ...961..198B} (B24) to stars with convective cores. We found best-fit models from a grid of tracks computed with MESA by fitting the observed frequencies, effective temperatures, and metallicities. Using each of these best-fit models, we performed structure inversions to compare the internal structure of the model to that of the star. These results, combined with the results of B24, show that our models reproduce the internal structure of around half of the stars examined. In cases where we find significant differences, we see an even split between models with dimensionless squared isothermal sound speed that is higher than the star and cases where it is too low. In contrast to the results of the stars with radiative cores, we did not find any significant correlations with the properties of our reference models.} \revv{We presented three models constructed with varying model physics as an example of the kinds of changes that could be tested in detail in future work using structure inversions. However, in our simple tests we found that the resulting changes to the model structure are much smaller than necessary to reproduce the structure inferred by our inversions.}

\revv{In both B24 and this work, the structure differences inferred by inversions remain unexplained. Particularly in light of the upcoming PLATO mission \citep{2014ExA....38..249R,2024arXiv240605447R}, it is important to continue improving our stellar models of these types of stars. \revvv{In addition to both improved forward modeling and detailed analysis of the glitches in the observed frequencies and frequency separation ratios}, structure inversions are an important part of ensuring that future models reproduce not only the global properties of the stars, but also their internal structure. To that end,} in future work we plan to test potential modifications to the physical ingredients in our stellar modeling using structure inversions. With these changes, we aim to consistently improve our models of the stars with significant differences without introducing discrepancies for the stars we currently model well.

\begin{acknowledgments}
The research leading to the presented results has received funding from the ERC Consolidator Grant DipolarSound (grant agreement \#101000296). SB acknowledges NSF grant AST-2205026.
This paper includes data collected by the Kepler mission. Funding for the Kepler mission is provided by the NASA Science Mission Directorate. 
In addition, this work has made use of data from the European Space Agency (ESA) mission {\it Gaia} (\url{https://www.cosmos.esa.int/gaia}), processed by the {\it Gaia}
Data Processing and Analysis Consortium (DPAC, \url{https://www.cosmos.esa.int/web/gaia/dpac/consortium}). Funding for the DPAC has been provided by national institutions, in particular the institutions
participating in the {\it Gaia} Multilateral Agreement.
We have also used the gaia-kepler.fun crossmatch database created by Megan Bedell.
\end{acknowledgments}



\appendix

\section{Appendix Modeling Details}
\subsection{Model Grids} 
\label{Appendix-grids} 
Here, we provide more details about the grid used to find our reference models. We use metal abundances scaled to the GS98 solar composition \citep{1998SSRv...85..161G}, and the corresponding high-temperature opacity tables from OPAL \citep{Iglesias1993, Iglesias1996}, and low-temperature opacity tables from \citet{Ferguson2005}. We blend the equation of state data from OPAL \citep{Rogers2002}, SCVH \citep{Saumon1995}, FreeEOS \citep{Irwin2004}, and Skye \citep{Jermyn2021} with the default settings. This blending is described in more detail in \citet{Jermyn2022}. Our nuclear reaction network is  \verb|pp_cno_extras_o18_ne22.net| and we use reaction rates from JINA REACLIB \citep{Cyburt2010} and  NACRE \citep{Angulo1999},  with additional tabulated weak reaction rates \citep{Fuller1985, Oda1994, Langanke2000}.  Electron screening is included via the prescription of \citet{Chugunov2007}. Thermal neutrino loss rates are from \citet{Itoh1996}. We use the time-dependent local convection formalism of \citet{1986A&A...160..116K}, which, as described in \citet{Jermyn2022}, refuses in the limit of long time steps to standard mixing length theory as described in \citet{1968pss..book.....C}. We use an Eddington-gray atmosphere and include the structure of the atmosphere out to an optical depth of \(\tau = 10^{-3}\) when calculating both our oscillation frequencies and structure kernels. 

\subsubsection{Convective Penetration}
As MESA does not implement convective penetration by default, we make use of the \verb|other_after_set_mixing_info| hook in \verb|run_star_extras|. This allows us to use MESA's procedure for calculating the extent of step overshooting region and simply change the temperature gradient after these regions have been identified. It also simplifies the process of including overshoot from the convective core and at the base of the outer convection zone. We include our routine for this at the Zenodo link provided in Section~\ref{Section-mods}. 

\subsection{Modeling Information} 
In Table~\ref{tab:obs}, we provide the non-seismic constraints used in our modeling procedure. For all but three stars, we take our spectroscopic measurements from \citet[][, Table 9]{Furlan_2018} and adopt their suggested uncertainties of 100K and 0.1 dex for \(T_{\rm{eff}}\) and [Fe/H], respectively. Two of our stars, KIC~434952 and KIC~5773345, are not analyzed by \citet{Furlan_2018}. In these cases, we use the values from \citet{Mathur_2017}. The values in \citet{Furlan_2018} for KIC~9965715 were found to be discrepant from other literature values, so instead we use the measurements from \citet{Morel_2021}.  To reduce the computation time when finding a best-fit model, we calculate \(\chi^2_{\rm{fit}}\) only for models within \(6\sigma\) of the observed effective temperature and metallicity and \(10\sigma\) of the observed luminosity. We primarily use the FLAME luminosity value from Gaia DR3  \citep{2016A&A...595A...1G, Gaia_DR3}, although some stars are only available in Gaia DR2 \citep{Gaia_DR2}, or not at all. 

\begin{figure} 
\gridline{\fig{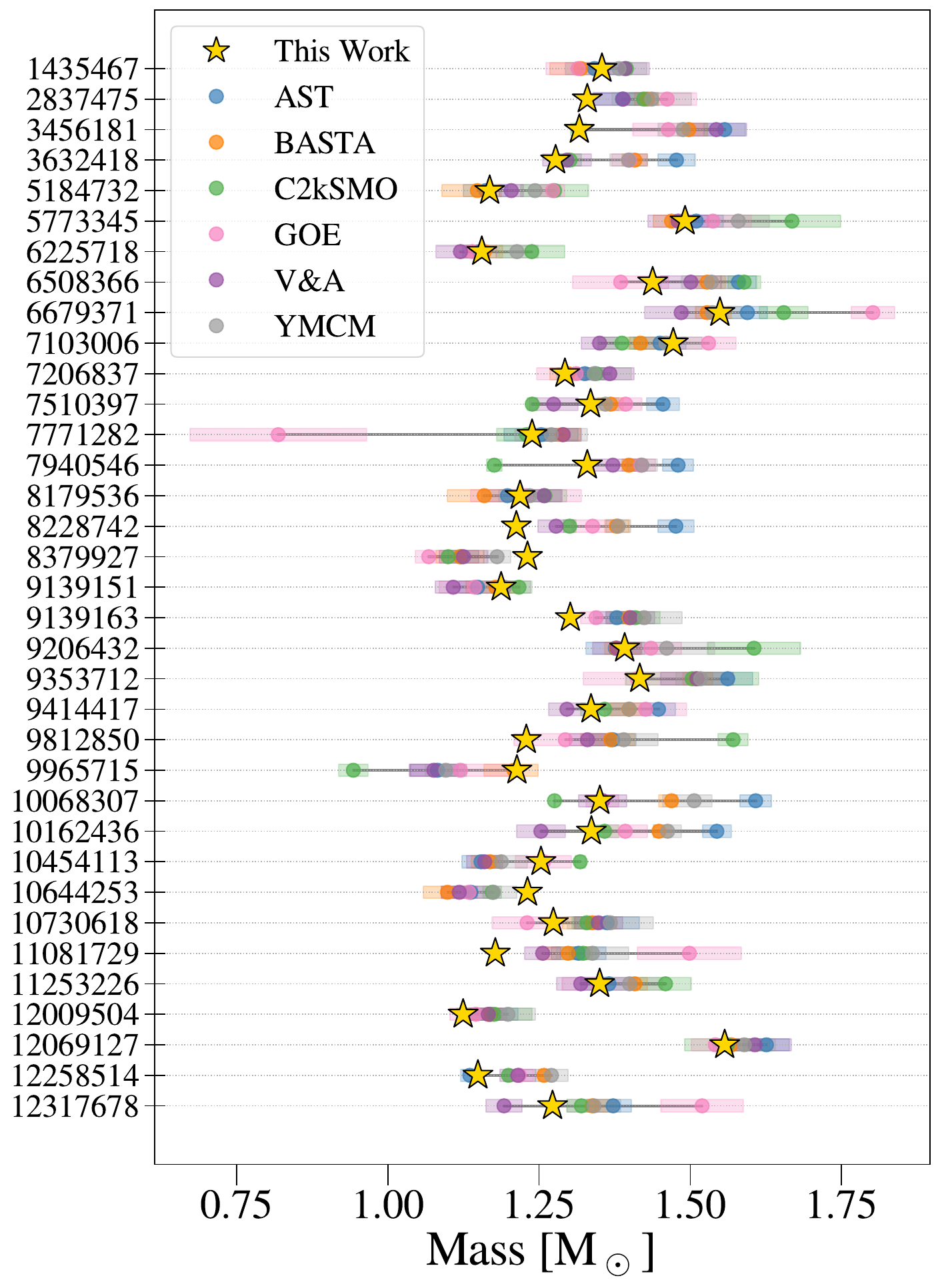}{0.4\textwidth}{}
          \fig{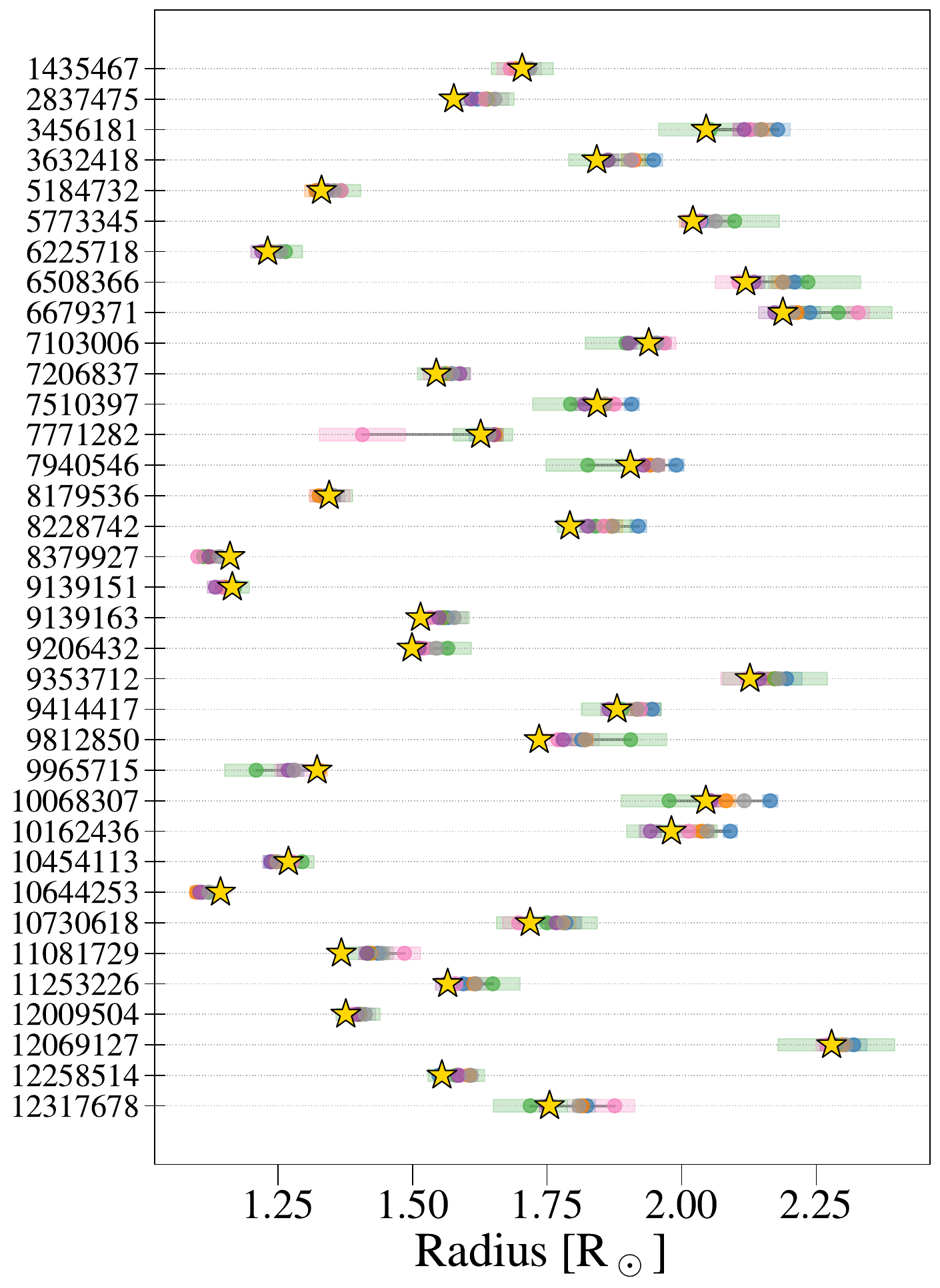}{0.4\textwidth}{}}
\gridline{\fig{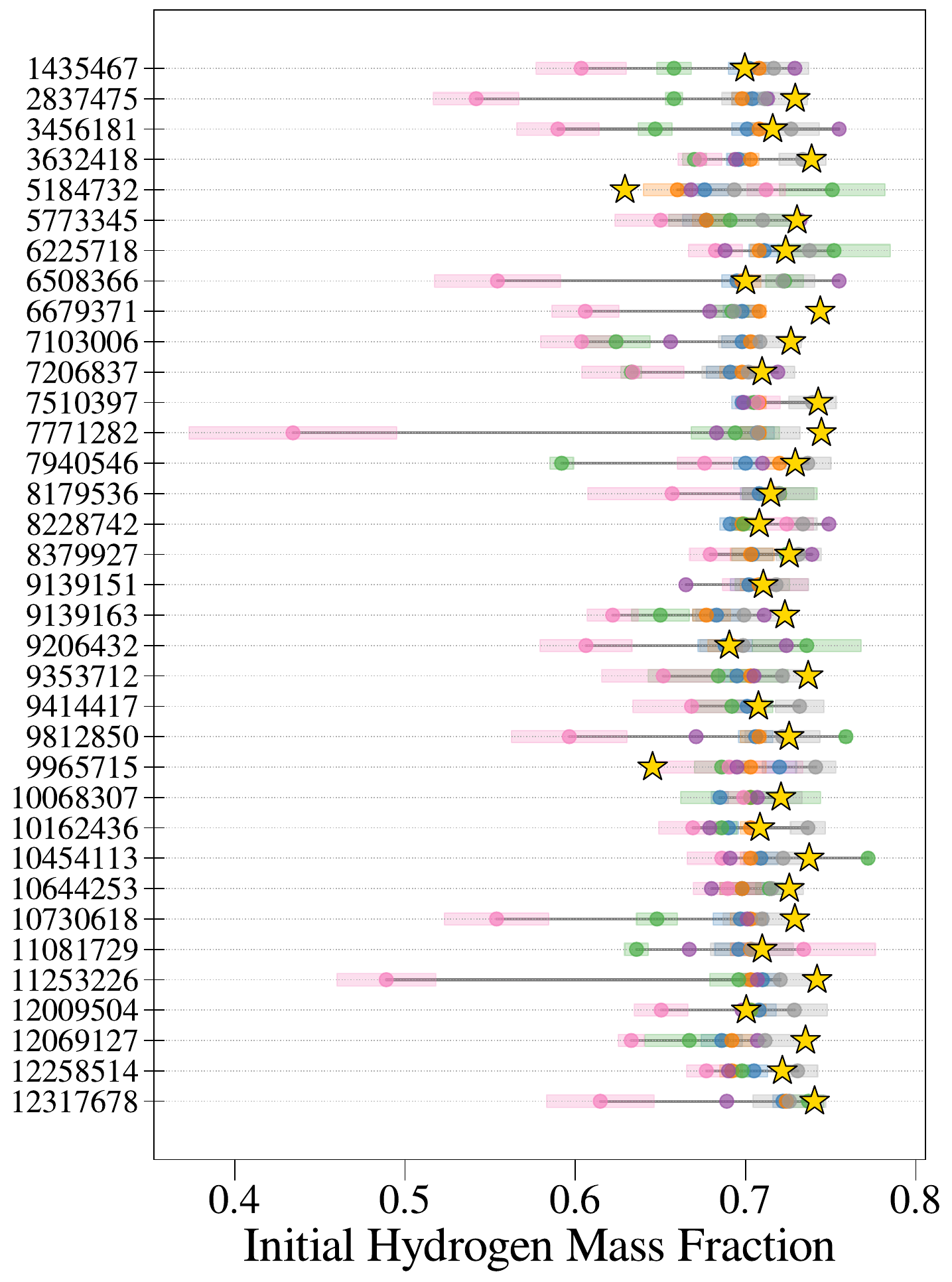}{0.4\textwidth}{}
          \fig{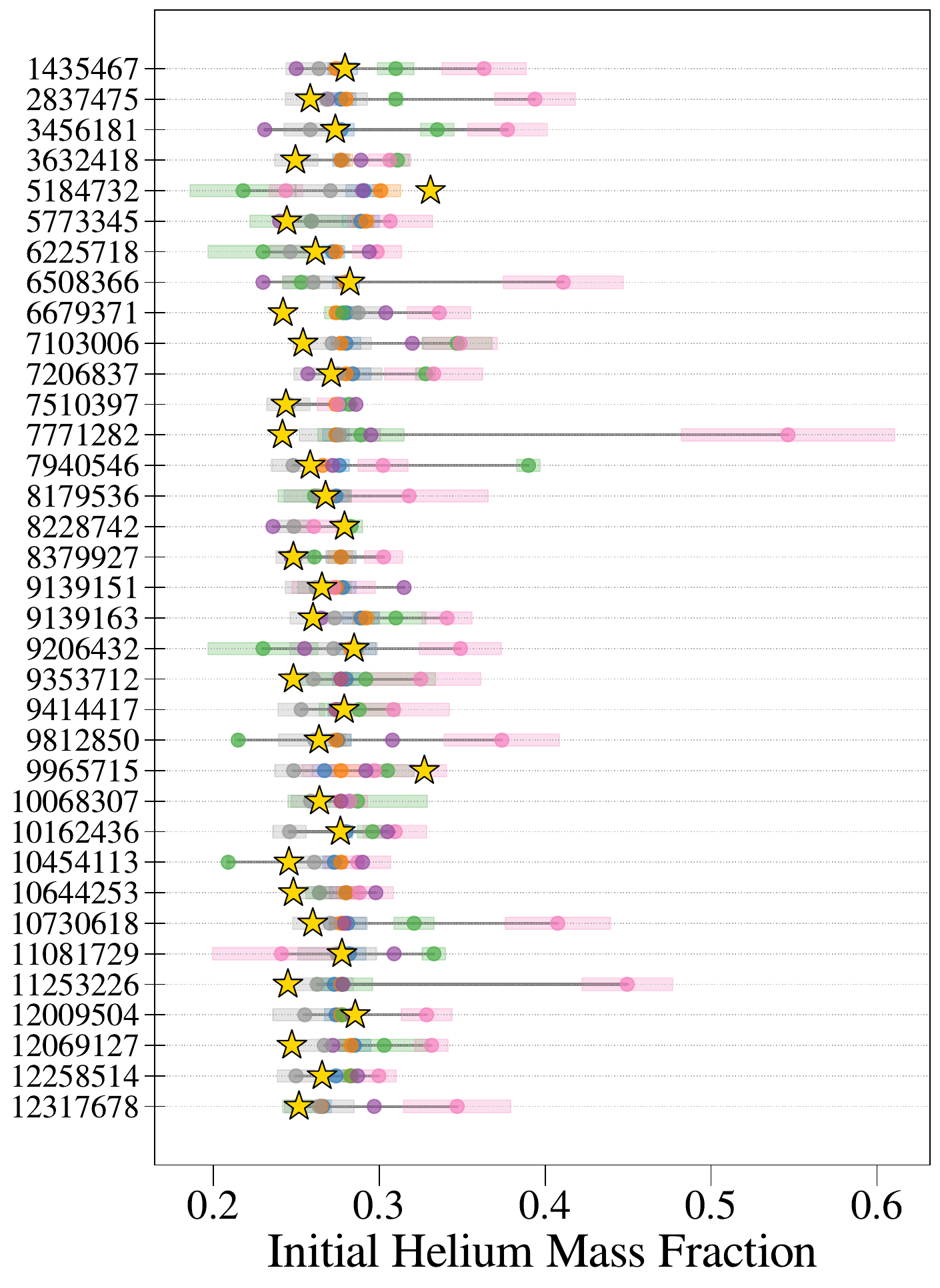}{0.4\textwidth}{}}
\caption{Comparison of our model parameters to several pipelines used in \citet{2017ApJ...835..173S}. The results of each pipeline are indicated with a dot and the uncertainties of that result a shaded region of the same color.} 
\label{fig:Leg_comp}
\end{figure}

\begin{figure} 
\gridline{\fig{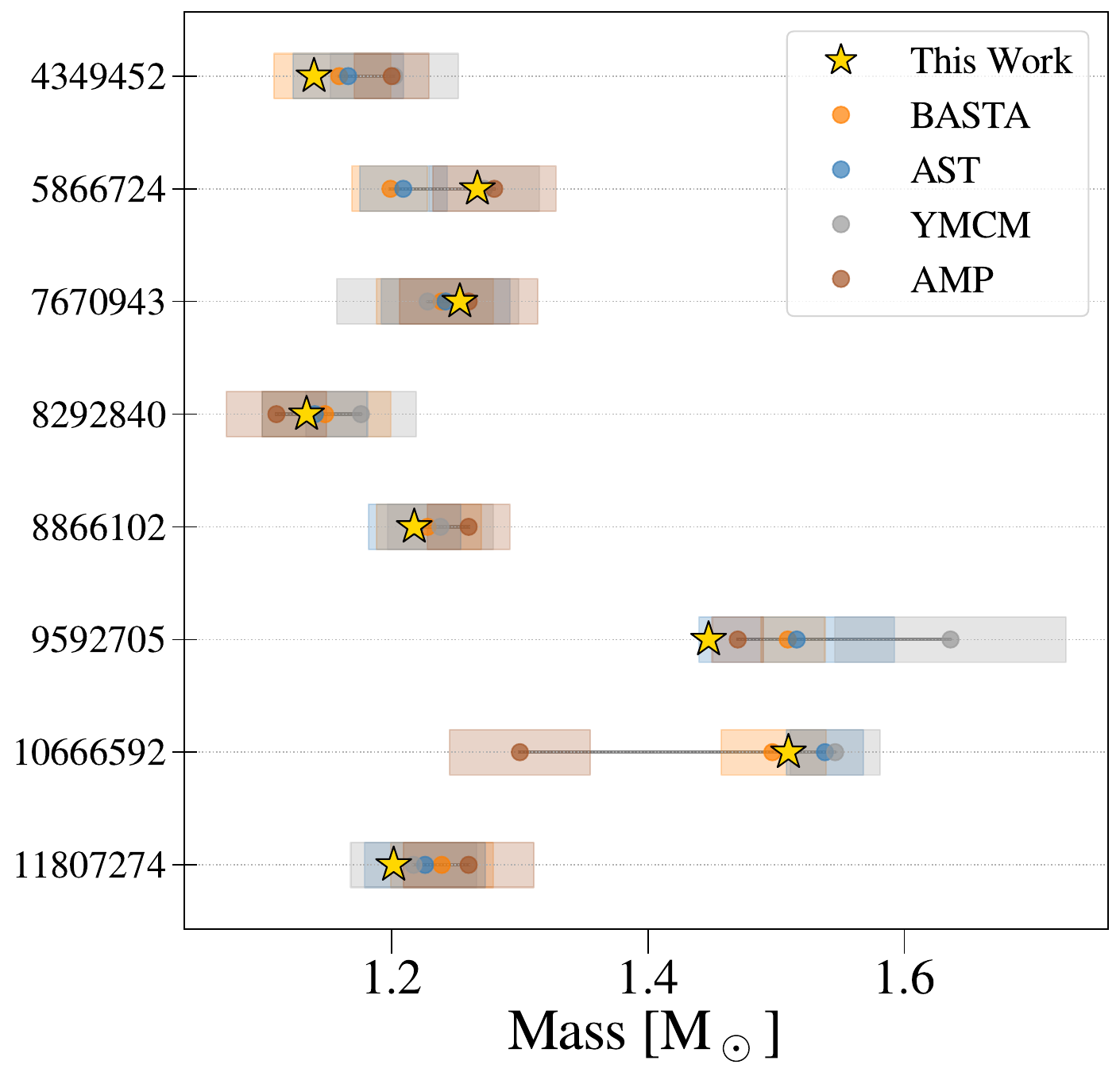}{0.4\textwidth}{}
          \fig{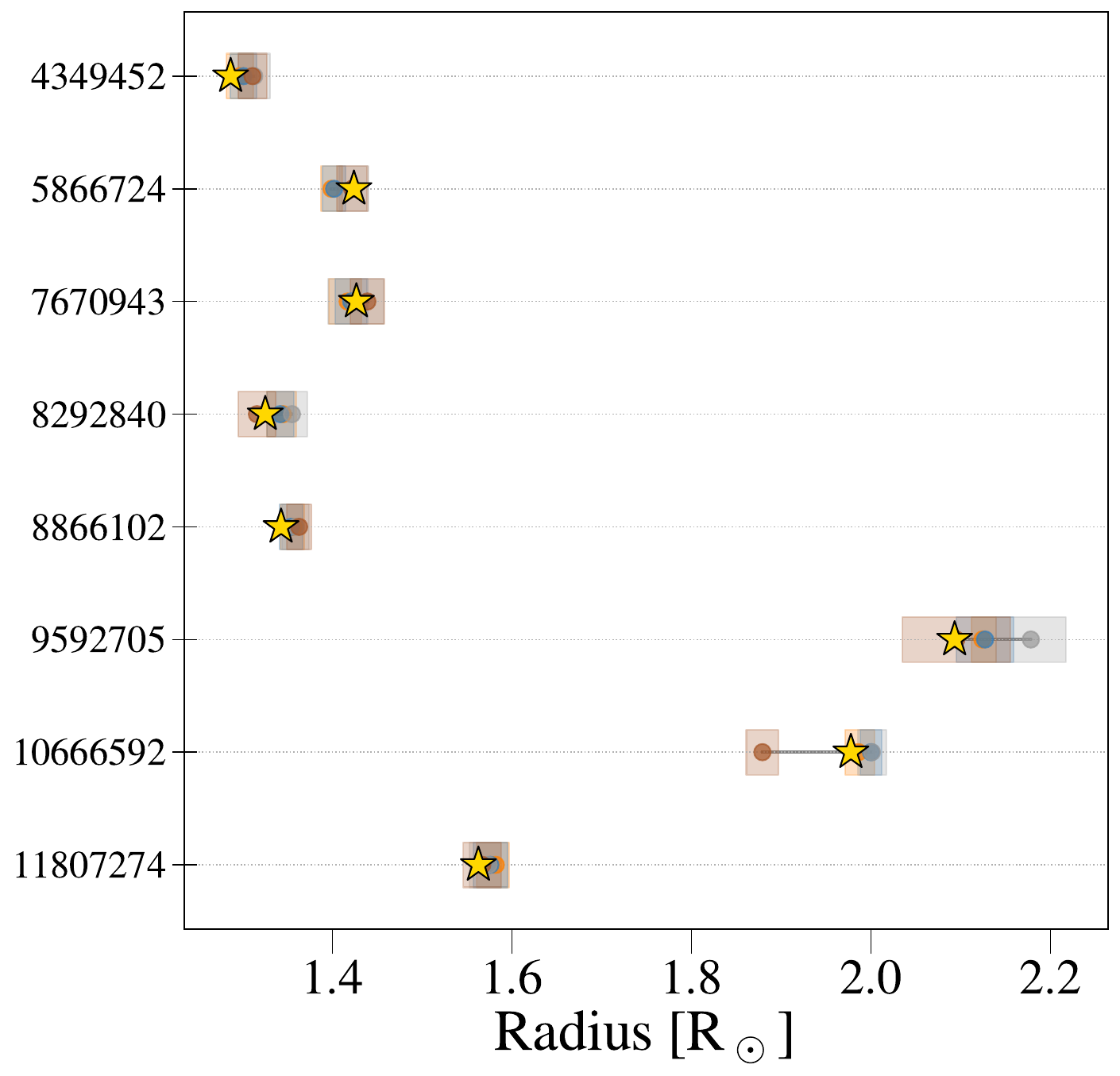}{0.4\textwidth}{}}
\gridline{\fig{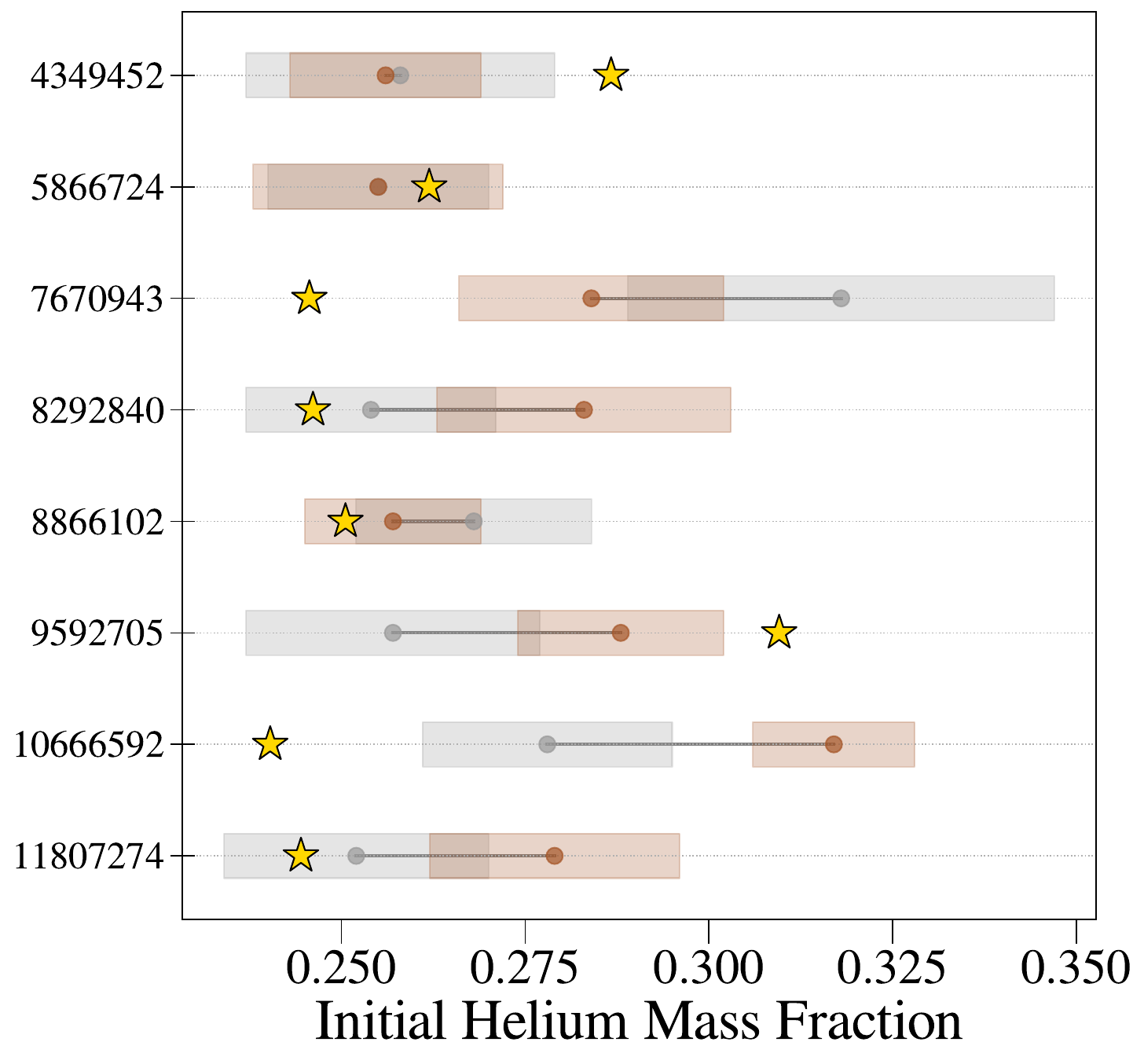}{0.4\textwidth}{}
          \fig{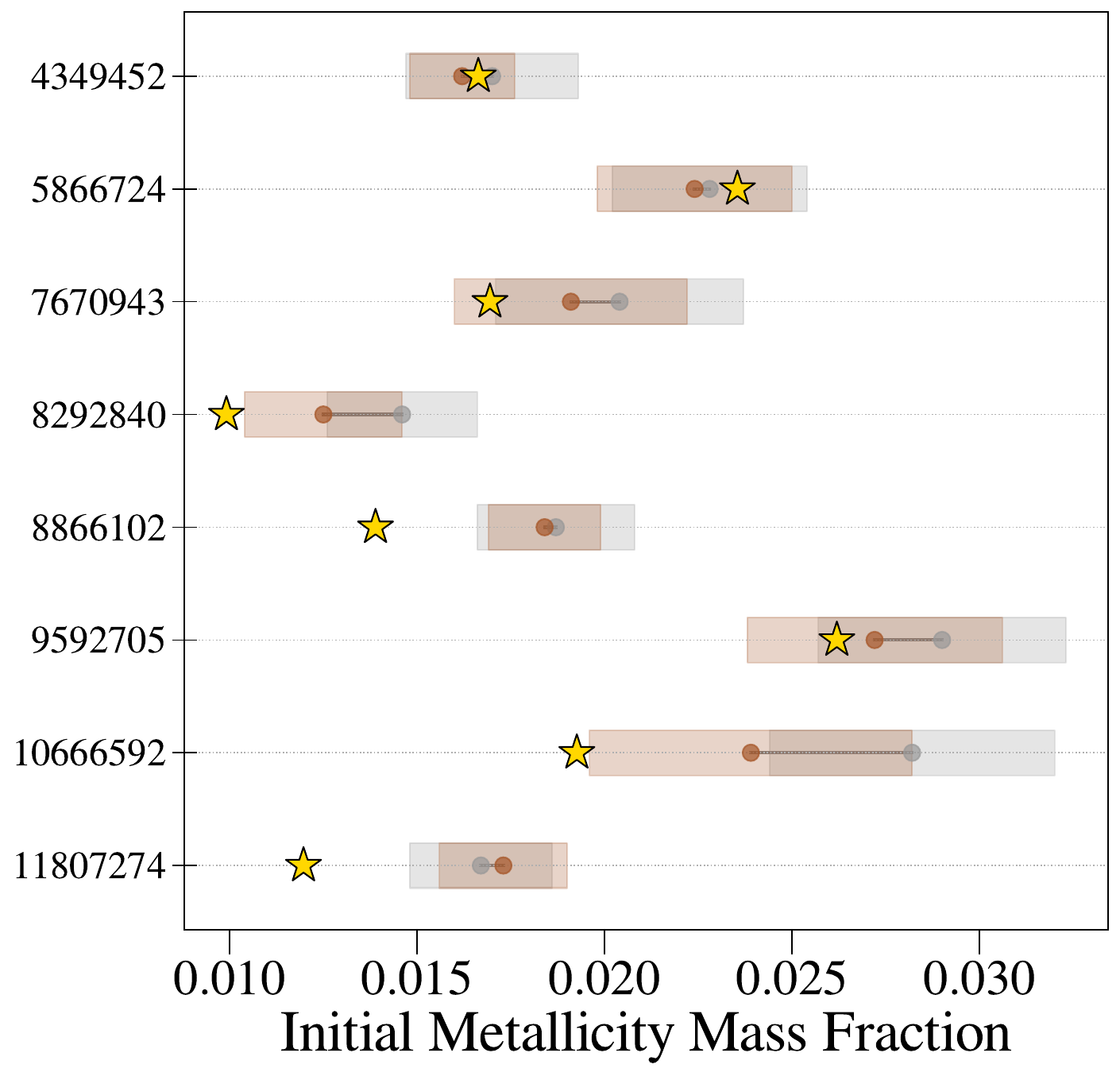}{0.4\textwidth}{}}
\caption{\revvv{Comparison of our model parameters to several pipelines used in \citet{2015MNRAS.452.2127S} The results of each pipeline are indicated with a dot and the uncertainties of that result a shaded region of the same color. The composition information is only available for the YMCM and AMP pipelines. } }
\label{fig:SA15_comp}
\end{figure}

Table~\ref{tab:RefModInfo} provides the parameters of our reference model for each star and the star's category based on our inversion results as defined in Section~\ref{sect:results}. \revvv{Note that these are the parameters of our reference model and should not be treated as inferred values of the stellar parameters.} \rev{We compare our \revvv{model} parameters to those reported in  \citet{2017ApJ...835..173S} in Figure~\ref{fig:Leg_comp}. \revvv{For the stars in our sample included in \citet{2015MNRAS.452.2127S} but not in \citet{2017ApJ...835..173S}, we compare our reference model parameters in Figure~\ref{fig:SA15_comp}}. In general, our values of mass and radius fall within the spread of values in \citet{2015MNRAS.452.2127S} and \citet{2017ApJ...835..173S} without any clear biases. \revv{This is in contrast to the composition of our models which show  a clear bias in favor of higher initial hydrogen mass fraction and lower initial helium mass fraction. We attribute this to our choice not to include diffusion in our models. Despite this bias, our values are still within the range of values predicted by the various pipelines used in \citet{2015MNRAS.452.2127S} and  \citet{2017ApJ...835..173S}.}}  

\begin{figure} 
\epsscale{0.5} 
\plotone{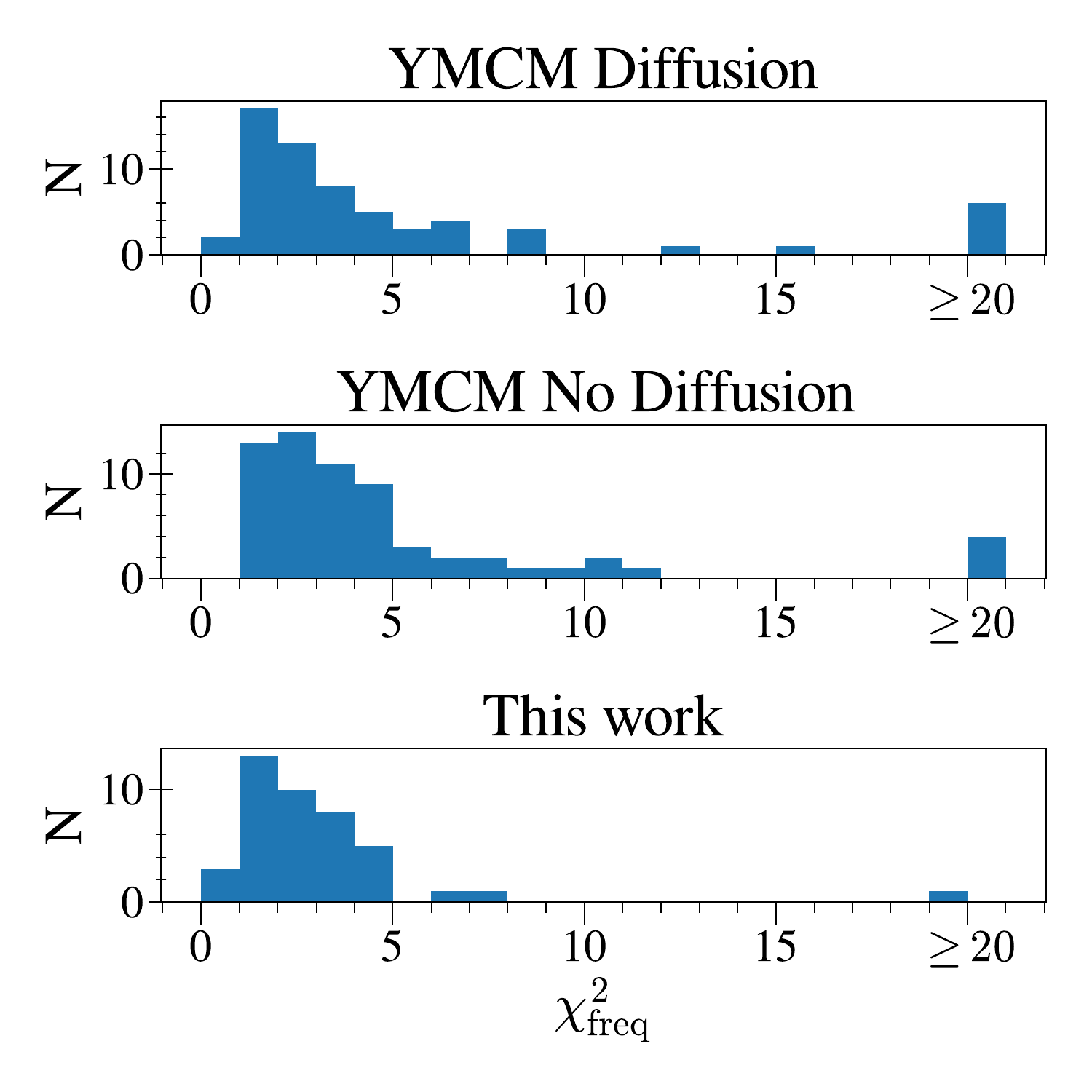} 
\caption{\revv{Distribution of $\chi^2_{\rm{fit}}$, as defined in Equation~\ref{equ:chi2_fit} for models fit using the YMCM pipeline of \citet{2015MNRAS.452.2127S, 2017ApJ...835..173S} and constructed with and without diffusion, as well as, the $\chi^2_{\rm{fit}}$ distribution of the models used in this work. Models with $\chi^2_{\rm{fit}} \geq 20$ have been collapsed into the final bin. }}
\label{fig:chi2_dist} 
\end{figure}

\revv{To compare the quality of our fits, we obtain the surface-term corrected frequencies of the best fitting models found using the YMCM pipeline of \citet{2015MNRAS.452.2127S, 2017ApJ...835..173S}. We obtained results for a set of models constructed both with and without diffusion. In Figure~\ref{fig:chi2_dist}, we plot the distributions of $\chi^2_{\rm{fit}}$ values for the YMCM models with and without diffusion and the reference models used for our inversions. We find that our overall distribution is similar, with fewer outliers resulting from our modeling procedure. }

\label{Appendix-mods} 

\begin{deluxetable}{lccc}
\tabletypesize{\small}
\tablecaption{Non-seismic observations} 
\label{tab:obs} 
\tablehead{\colhead{Star} & \colhead{\(T_{\rm{eff}}\) [K]} & \colhead{[Fe/H]} & \colhead{\(L \: [L_\odot]\)}}  

\startdata
KIC 1435467 & 6325.0$\pm$100.0\tablenotemark{a} &   0.04$\pm$0.1\tablenotemark{a} & 4.051$\pm$0.073\tablenotemark{d} \\
 KIC 2837475 & 6488.0$\pm$100.0\tablenotemark{a} &  -0.07$\pm$0.1\tablenotemark{a} &   4.7$\pm$0.019\tablenotemark{d} \\
 KIC 3456181 & 6214.0$\pm$100.0\tablenotemark{a} &  -0.26$\pm$0.1\tablenotemark{a} &   6.72$\pm$0.04\tablenotemark{d} \\
 KIC 3632418 & 6112.0$\pm$100.0\tablenotemark{a} &  -0.16$\pm$0.1\tablenotemark{a} & 4.973$\pm$0.073\tablenotemark{d} \\
 KIC 4349452 &  6267.0$\pm$81.0\tablenotemark{b} & -0.06$\pm$0.15\tablenotemark{b} & 2.379$\pm$0.015\tablenotemark{d} \\
 KIC 5184732 & 5874.0$\pm$100.0\tablenotemark{a} &   0.41$\pm$0.1\tablenotemark{a} & 1.995$\pm$0.008\tablenotemark{d} \\
 KIC 5773345 &  6127.0$\pm$82.0\tablenotemark{b} &   0.21$\pm$0.1\tablenotemark{b} &  5.429$\pm$0.03\tablenotemark{d} \\
 KIC 5866724 & 6138.0$\pm$100.0\tablenotemark{a} &   0.14$\pm$0.1\tablenotemark{a} & 2.667$\pm$0.017\tablenotemark{d} \\
 KIC 6225718 & 6203.0$\pm$100.0\tablenotemark{a} &  -0.12$\pm$0.1\tablenotemark{a} & 2.208$\pm$0.007\tablenotemark{d} \\
 KIC 6508366 & 6249.0$\pm$100.0\tablenotemark{a} &  -0.06$\pm$0.1\tablenotemark{a} & 6.959$\pm$0.031\tablenotemark{d} \\
 KIC 6679371 & 6387.0$\pm$100.0\tablenotemark{a} &  -0.04$\pm$0.1\tablenotemark{a} & 7.865$\pm$0.036\tablenotemark{d} \\
 KIC 7103006 & 6362.0$\pm$100.0\tablenotemark{a} &   0.05$\pm$0.1\tablenotemark{a} & 5.747$\pm$0.019\tablenotemark{d} \\
 KIC 7206837 & 6325.0$\pm$100.0\tablenotemark{a} &   0.12$\pm$0.1\tablenotemark{a} & 3.664$\pm$0.022\tablenotemark{d} \\
 KIC 7510397 & 6109.0$\pm$100.0\tablenotemark{a} &  -0.25$\pm$0.1\tablenotemark{a} &      \nodata \\
 KIC 7670943 & 6302.0$\pm$100.0\tablenotemark{a} &   0.01$\pm$0.1\tablenotemark{a} &  2.98$\pm$0.041\tablenotemark{d} \\
 KIC 7771282 & 6138.0$\pm$100.0\tablenotemark{a} &  -0.07$\pm$0.1\tablenotemark{a} & 3.654$\pm$0.029\tablenotemark{d} \\
 KIC 7940546 & 6126.0$\pm$100.0\tablenotemark{a} &  -0.27$\pm$0.1\tablenotemark{a} & 5.443$\pm$0.059\tablenotemark{d} \\
 KIC 8179536 & 6281.0$\pm$100.0\tablenotemark{a} &  -0.04$\pm$0.1\tablenotemark{a} & 2.666$\pm$0.015\tablenotemark{d} \\
 KIC 8228742 & 6046.0$\pm$100.0\tablenotemark{a} &  -0.09$\pm$0.1\tablenotemark{a} & 4.273$\pm$0.042\tablenotemark{d} \\
 KIC 8292840 & 6212.0$\pm$100.0\tablenotemark{a} &  -0.21$\pm$0.1\tablenotemark{a} & 2.608$\pm$0.054\tablenotemark{d} \\
 KIC 8379927 &  6022.0$\pm$77.0\tablenotemark{b} & -0.24$\pm$0.35\tablenotemark{b} &     \nodata \\
 KIC 8866102 & 6273.0$\pm$100.0\tablenotemark{a} &  -0.09$\pm$0.1\tablenotemark{a} & 2.814$\pm$0.013\tablenotemark{d} \\
 KIC 9139151 & 6040.0$\pm$100.0\tablenotemark{a} &   0.04$\pm$0.1\tablenotemark{a} & 1.669$\pm$0.007\tablenotemark{d} \\
 KIC 9139163 & 6350.0$\pm$100.0\tablenotemark{a} &   0.09$\pm$0.1\tablenotemark{a} & 3.755$\pm$0.028\tablenotemark{d} \\
 KIC 9206432 & 6490.0$\pm$100.0\tablenotemark{a} &   0.17$\pm$0.1\tablenotemark{a} & 3.934$\pm$0.029\tablenotemark{d} \\
 KIC 9353712 & 6140.0$\pm$100.0\tablenotemark{a} &  -0.09$\pm$0.1\tablenotemark{a} & 6.346$\pm$0.057\tablenotemark{d} \\
 KIC 9414417 & 6283.0$\pm$100.0\tablenotemark{a} &  -0.09$\pm$0.1\tablenotemark{a} & 5.502$\pm$0.024\tablenotemark{d} \\
 KIC 9592705 & 6148.0$\pm$100.0\tablenotemark{a} &   0.27$\pm$0.1\tablenotemark{a} & 5.987$\pm$0.098\tablenotemark{d} \\
 KIC 9812850 & 6314.0$\pm$100.0\tablenotemark{a} &  -0.18$\pm$0.1\tablenotemark{a} & 4.621$\pm$0.021\tablenotemark{d} \\
 KIC 9965715 &  6335.0$\pm$40.0\tablenotemark{c} &  0.29$\pm$0.04\tablenotemark{c} & 2.716$\pm$0.042\tablenotemark{d} \\
KIC 10068307 & 6050.0$\pm$100.0\tablenotemark{a} &  -0.21$\pm$0.1\tablenotemark{a} & 5.391$\pm$0.021\tablenotemark{d} \\
KIC 10162436 & 6134.0$\pm$100.0\tablenotemark{a} &  -0.14$\pm$0.1\tablenotemark{a} & 5.374$\pm$0.019\tablenotemark{d} \\
KIC 10454113 & 6136.0$\pm$100.0\tablenotemark{a} &  -0.07$\pm$0.1\tablenotemark{a} & 2.784$\pm$0.046\tablenotemark{d} \\
KIC 10644253 & 6020.0$\pm$100.0\tablenotemark{a} &   0.09$\pm$0.1\tablenotemark{a} & 1.515$\pm$0.006\tablenotemark{d} \\
KIC 10666592 & 6264.0$\pm$100.0\tablenotemark{a} &   0.01$\pm$0.1\tablenotemark{a} & 6.183$\pm$0.081\tablenotemark{e} \\
KIC 10730618 & 6423.0$\pm$168.0\tablenotemark{b} &  -0.16$\pm$0.3\tablenotemark{b} &  4.545$\pm$0.04\tablenotemark{d} \\
KIC 11081729 & 6416.0$\pm$100.0\tablenotemark{a} &  -0.13$\pm$0.1\tablenotemark{a} & 3.386$\pm$0.054\tablenotemark{d} \\
KIC 11253226 & 6474.0$\pm$100.0\tablenotemark{a} &  -0.19$\pm$0.1\tablenotemark{a} & 4.605$\pm$0.032\tablenotemark{d} \\
KIC 11807274 & 6150.0$\pm$100.0\tablenotemark{a} &  -0.12$\pm$0.1\tablenotemark{a} &  3.34$\pm$0.027\tablenotemark{d} \\
KIC 12009504 & 6129.0$\pm$100.0\tablenotemark{a} &  -0.08$\pm$0.1\tablenotemark{a} & 2.659$\pm$0.009\tablenotemark{d} \\
KIC 12069127 & 6186.0$\pm$100.0\tablenotemark{a} &   0.03$\pm$0.1\tablenotemark{a} & 7.677$\pm$0.082\tablenotemark{d} \\
KIC 12258514 & 5948.0$\pm$100.0\tablenotemark{a} &   0.01$\pm$0.1\tablenotemark{a} & 3.016$\pm$0.009\tablenotemark{d} \\
KIC 12317678 & 6395.0$\pm$100.0\tablenotemark{a} &  -0.42$\pm$0.1\tablenotemark{a} & 5.653$\pm$0.091\tablenotemark{e} \\
\enddata
\tablerefs{(a) \cite{Furlan_2018}, (b) \cite{Mathur_2017}, (c) \cite{Morel_2021}, (d) \cite{Gaia_DR3}, (e) \cite{Gaia_DR2}}
\end{deluxetable}

\begin{deluxetable}{lccccccccc}
\tablecaption{Reference Model Parameters} 
\label{tab:RefModInfo} 
\tablehead{\colhead{Star} & \colhead{$M [\rm{M}_{\odot}]$} & \colhead{$Y_{\rm{initial}}$} & \colhead{$Z_{\rm{initial}}$} & \colhead{$\alpha_{\rm{mlt}}$} & \colhead{$f_{ov}$} & \colhead{$X_{c}$} & \colhead{Age [Gyr]} &\colhead{$\chi^2_{\rm{fit}}$} & \colhead{Inversion Results Category}}

\startdata
 KIC 1435467 & 1.3540 & 0.2794 & 0.0209 & 2.2396 & 0.0231 & 0.2689 & 2.6253 &  2.32 &  H \\
 KIC 2837475 & 1.3297 & 0.2584 & 0.0123 & 2.1497 & 0.0243 & 0.4000 & 2.0137 &   7.2 &  A \\
 KIC 3456181 & 1.3165 & 0.2736 & 0.0104 & 2.2254 & 0.0504 & 0.2191 & 2.9891 &   6.1 &  H \\
 KIC 3632418 & 1.2773 & 0.2496 & 0.0116 & 1.9093 & 0.0219 & 0.1037 & 3.6243 &  2.29 &  A \\
 KIC 4349452 & 1.1394 & 0.2867 & 0.0166 & 2.0684 & 0.0344 & 0.3853 & 3.0303 &  1.13 &  H \\
 KIC 5184732 & 1.1683 & 0.3309 & 0.0399 & 2.2366 & 0.0124 & 0.1589 & 4.3569 & 27.77 &  A \\
 KIC 5773345 & 1.4914 & 0.2443 & 0.0254 & 2.0132 & 0.0464 & 0.2997 & 3.1090 &   3.8 &  A \\
 KIC 5866724 & 1.2668 & 0.2619 & 0.0235 & 2.1377 & 0.0250 & 0.3511 & 3.2562 &  1.82 &  L \\
 KIC 6225718 & 1.1553 & 0.2617 & 0.0147 & 2.3102 & 0.0392 & 0.4493 & 2.7897 &  7.73 & HL \\
 KIC 6508366 & 1.4378 & 0.2824 & 0.0175 & 2.1715 & 0.0394 & 0.2403 & 2.3956 &  5.88 &  L \\
 KIC 6679371 & 1.5490 & 0.2421 & 0.0140 & 2.1485 & 0.0110 & 0.1463 & 1.9634 &  4.38 &  A \\
 KIC 7103006 & 1.4718 & 0.2541 & 0.0192 & 2.1252 & 0.0294 & 0.3029 & 2.3479 &  1.49 &  H \\
 KIC 7206837 & 1.2928 & 0.2711 & 0.0191 & 1.9863 & 0.0355 & 0.4079 & 2.6459 &  2.27 &  A \\
 KIC 7510397 & 1.3352 & 0.2438 & 0.0134 & 2.1562 & 0.0168 & 0.0838 & 3.3145 &  5.46 &  A \\
 KIC 7670943 & 1.2531 & 0.2456 & 0.0169 & 2.3065 & 0.0165 & 0.2911 & 3.3181 &  1.62 &  A \\
 KIC 7771282 & 1.2384 & 0.2417 & 0.0137 & 2.0294 & 0.0401 & 0.2714 & 4.2713 &  1.85 &  L \\
 KIC 7940546 & 1.3297 & 0.2584 & 0.0123 & 2.1497 & 0.0243 & 0.1338 & 3.0299 &  7.97 &  H \\
 KIC 8179536 & 1.2186 & 0.2677 & 0.0174 & 2.1525 & 0.0329 & 0.4253 & 2.5597 &   3.1 &  H \\
 KIC 8228742 & 1.2124 & 0.2791 & 0.0128 & 2.1034 & 0.0261 & 0.0278 & 3.9881 &  3.35 &  A \\
 KIC 8292840 & 1.1336 & 0.2461 & 0.0099 & 1.9458 & 0.0127 & 0.1372 & 3.6114 &  2.28 &  H \\
 KIC 8379927 & 1.2308 & 0.2483 & 0.0259 & 2.1166 & 0.0085 & 0.5646 & 1.3710 &  6.51 &  L \\
 KIC 8866102 & 1.2175 & 0.2505 & 0.0139 & 2.1005 & 0.0026 & 0.2313 & 2.4654 &  2.06 &  A \\
 KIC 9139151 & 1.1872 & 0.2655 & 0.0240 & 2.3548 & 0.0143 & 0.4026 & 2.3697 &  4.92 &  L \\
 KIC 9139163 & 1.3016 & 0.2601 & 0.0167 & 2.0525 & 0.0370 & 0.4446 & 2.3699 &  5.65 & LH \\
 KIC 9206432 & 1.3915 & 0.2849 & 0.0245 & 2.0034 & 0.0062 & 0.5158 & 1.0102 &  4.49 &  A \\
 KIC 9353712 & 1.4165 & 0.2483 & 0.0149 & 2.0718 & 0.0390 & 0.2046 & 3.0382 &  1.94 &  A \\
 KIC 9414417 & 1.3359 & 0.2789 & 0.0135 & 2.3704 & 0.0200 & 0.1195 & 2.6318 &  2.79 &  A \\
 KIC 9592705 & 1.4472 & 0.3096 & 0.0262 & 2.0641 & 0.0135 & 0.0848 & 2.3208 &  3.29 &  H \\
 KIC 9812850 & 1.2289 & 0.2637 & 0.0106 & 2.1612 & 0.0590 & 0.2986 & 3.7670 &  1.27 &  H \\
 KIC 9965715 & 1.2133 & 0.3272 & 0.0273 & 1.9746 & 0.0147 & 0.4504 & 1.5278 & 20.35 &  A \\
KIC 10068307 & 1.3505 & 0.2640 & 0.0151 & 2.2139 & 0.0354 & 0.0988 & 3.4016 &  7.49 &  A \\
KIC 10162436 & 1.3365 & 0.2766 & 0.0149 & 2.1053 & 0.0276 & 0.1281 & 3.0031 &  3.87 &  A \\
KIC 10454113 & 1.2531 & 0.2456 & 0.0169 & 2.3065 & 0.0165 & 0.4990 & 1.6690 &  9.16 &  A \\
KIC 10644253 & 1.2308 & 0.2483 & 0.0259 & 2.1166 & 0.0085 & 0.6138 & 0.9882 &   2.5 &  H \\
KIC 10666592 & 1.5095 & 0.2403 & 0.0193 & 2.0247 & 0.0130 & 0.2321 & 2.2907 &  1.86 &  A \\
KIC 10730618 & 1.2735 & 0.2600 & 0.0110 & 2.2181 & 0.0435 & 0.2946 & 3.1335 &  3.46 &  A \\
KIC 11081729 & 1.1776 & 0.2775 & 0.0126 & 1.9915 & 0.0186 & 0.3516 & 2.4767 &  4.65 &  L \\
KIC 11253226 & 1.3502 & 0.2450 & 0.0129 & 2.0507 & 0.0410 & 0.4803 & 2.0173 &   7.3 &  A \\
KIC 11807274 & 1.2015 & 0.2445 & 0.0120 & 2.1044 & 0.0203 & 0.1399 & 4.3142 &  3.65 &  H \\
KIC 12009504 & 1.1244 & 0.2856 & 0.0140 & 2.1580 & 0.0232 & 0.2365 & 3.8576 &   4.7 &  A \\
KIC 12069127 & 1.5569 & 0.2474 & 0.0173 & 2.0223 & 0.0103 & 0.0847 & 2.1788 &   1.8 &  A \\
KIC 12258514 & 1.1489 & 0.2656 & 0.0127 & 2.1986 & 0.0307 & 0.0890 & 4.9999 & 10.74 &  A \\
KIC 12317678 & 1.2723 & 0.2517 & 0.0077 & 1.9616 & 0.0089 & 0.0771 & 2.7866 &   4.0 &  A \\
\enddata
\end{deluxetable}

\section{All inversion results} 
\label{Appendix-inv} 
\begin{figure} 

    \epsscale{1}
    \plotone{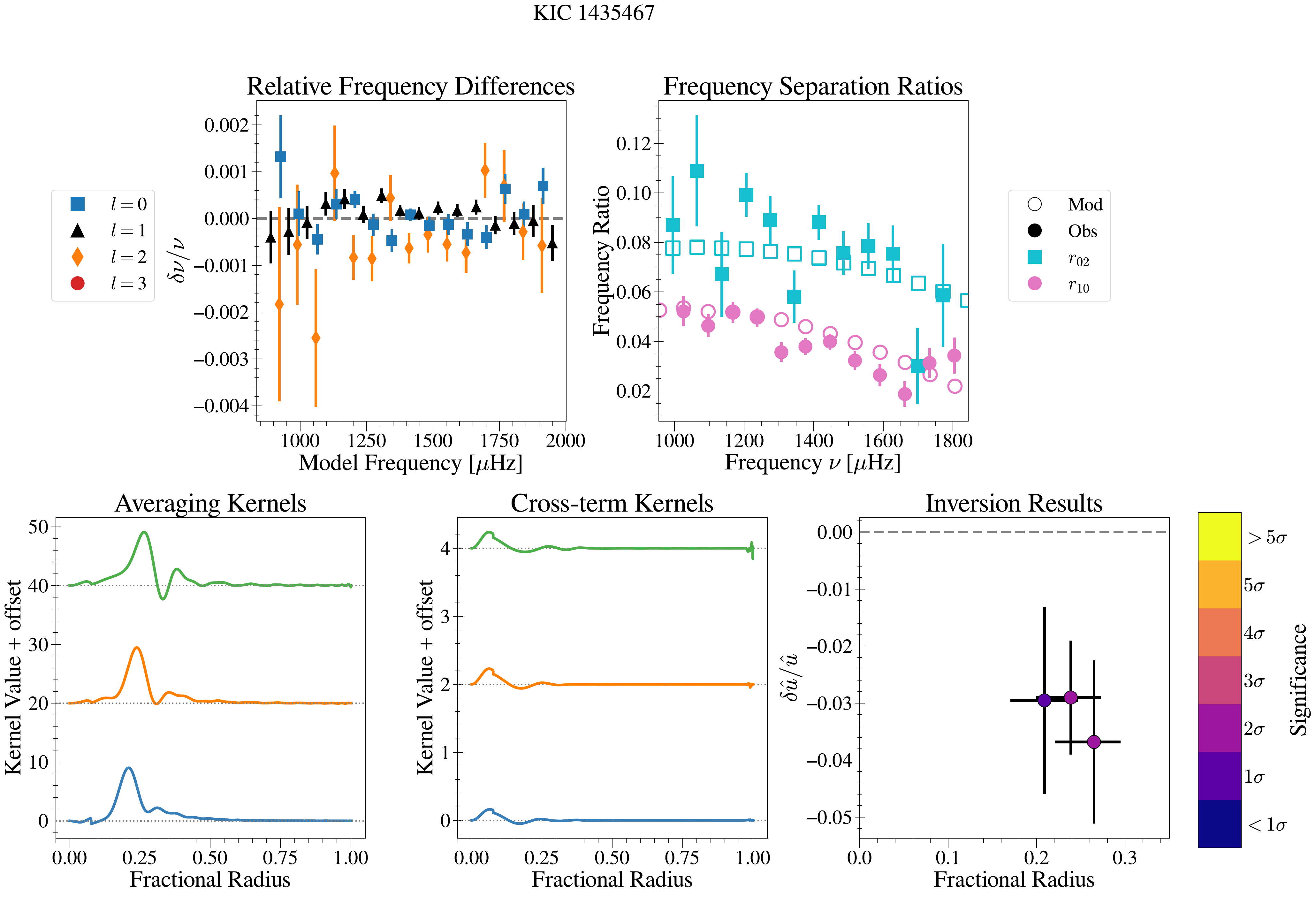}
    \caption{Modeling and inversion results for KIC 1435467.The top row shows the relative frequency differences after correcting for surface effects (left) and the frequency separation ratios (right). \rev{} The bottom row plots the averaging kernels (left), cross-term kernels (center), and inversion results (right). Note that the $y$-axis scale differs between the left and center plots. The complete figure set (43 images) is available in the online journal.}
    \label{fig:full_summary} 
\end{figure}

We provide plots that show the frequency differences, separation ratios, averaging kernels, cross-term kernels, and inversion results. In Figure~\ref{fig:full_summary} we show as an example KIC~1435467 with the results of all stars available in the online journal. 

\figsetstart
\figsetnum{\ref{fig:full_summary}}
\figsettitle{Modeling and Inversion Results}
\figsetgrpstart
\figsetgrpstart
\figsetgrpnum{\ref{fig:full_summary}.23}
\figsetgrptitle{KIC1435467}
\figsetplot{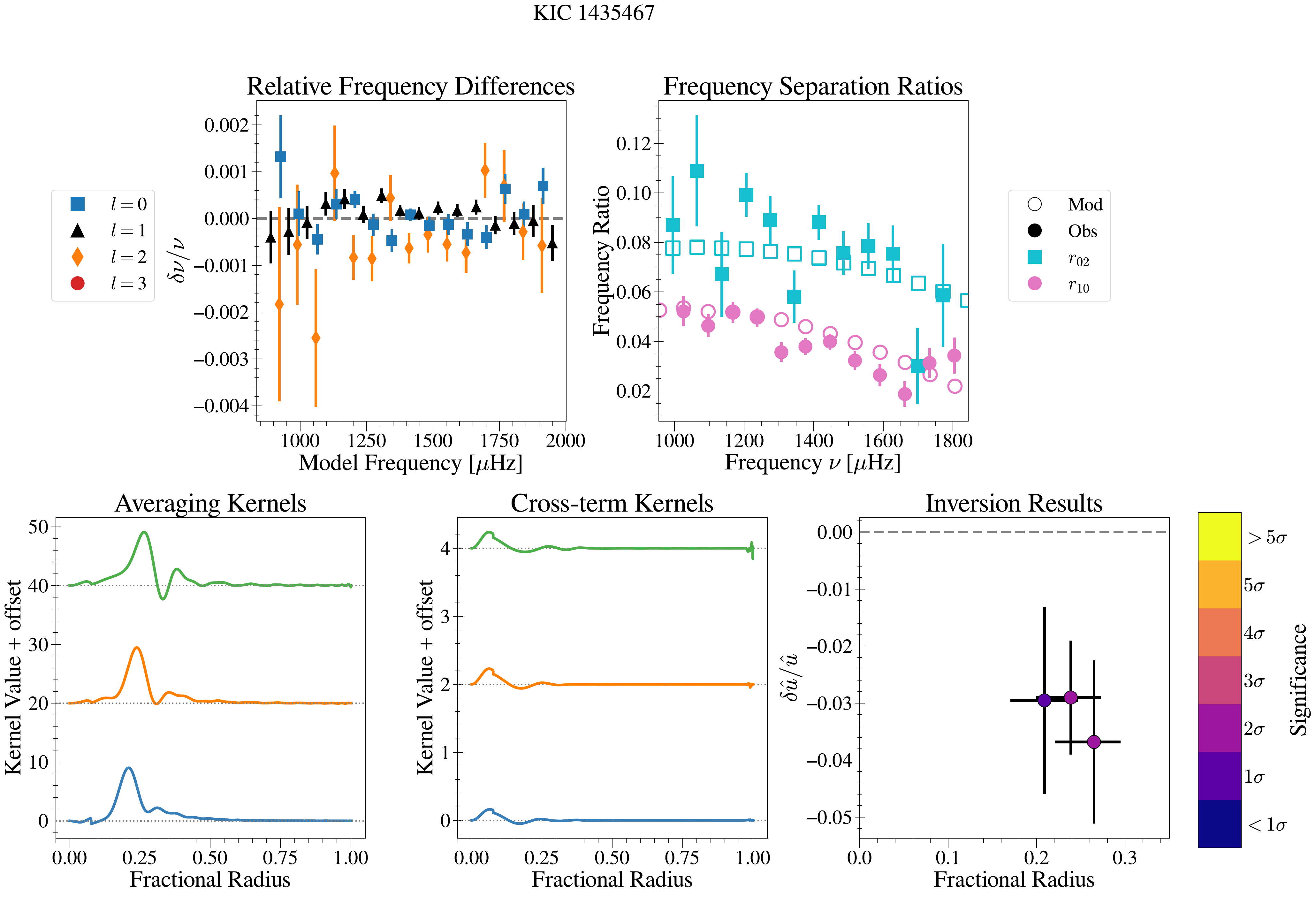}
\figsetgrpnote{Modeling and inversion results for KIC 1435467.The top row shows the relative frequency differences after correcting for surface effects (left) and the frequency separation ratios (right).  The bottom row plots the averaging kernels (left), cross-term kernels (center), and inversion results (right). Note that the $y$-axis scale differs between the left and center plots.}
\figsetgrpend
\figsetend

\clearpage

\bibliography{cc_inv_Buchele}{}
\bibliographystyle{aasjournal}



\end{document}